\documentclass[11pt]{article}
\usepackage{amsmath,amssymb,graphicx,epsfig}
\usepackage{fleqn} 

\def \lsim{\mathrel{\vcenter
{\hbox{$<$}\nointerlineskip\hbox{$\sim$}}}}

\def\bari{i \hspace{-3pt} \bar{\;\raisebox{3.7pt}{}}}
\def\barj{j \hspace{-3pt} \bar{\;\raisebox{3.7pt}{}}}
\def\alphacap{{\mbox{$\alpha$}}}
\def\betacap{{\mbox{$\beta$}}}

\def\bea{\begin{eqnarray}}
\def\eea{\end{eqnarray}}
\def\be{\begin{equation}}
\def\ee{\end{equation}}
\def\ba{\begin{array}}
\def\ea{\end{array}}
\def\nn{\nonumber}
\def\a{& \hspace{-11pt}}

\font\tenrsfs=rsfs10
\font\sevenrsfs=rsfs7
\font\fiversfs=rsfs5
\newfam\rsfsfam
\textfont\rsfsfam=\tenrsfs
\scriptfont\rsfsfam=\sevenrsfs
\scriptscriptfont\rsfsfam=\fiversfs
\def\mathscr#1{{\fam\rsfsfam\relax#1}}

\begin{document}

\thispagestyle{empty}

\begin{center}

$\;$

\vspace{1cm}

{\huge \bf Soft masses in superstring models \\[2mm] with anomalous $U(1)$ symmetries}

\vspace{1.5cm}

{\Large \bf Claudio~A.~Scrucca} \\[2mm] 

\vspace{0.8cm}

{\large \em Institut de Th\'eorie des Ph\'enom\`enes Physiques\\ 
Ecole Polytechnique F\'ed\'erale de Lausanne\\ 
CH-1015 Lausanne, Switzerland\\}
{\tt E-mail:~claudio.scrucca@epfl.ch}

\vspace{0.2cm}

\end{center}

\vspace{1cm}

\centerline{\bf \large Abstract}
\begin{quote}

We analyze the general structure of soft scalar masses emerging in superstring models involving 
anomalous $U(1)$ symmetries, with the aim of characterizing more systematically the 
circumstances under which they can happen to be flavor universal. We consider both 
heterotic orbifold and intersecting brane models, possibly with several anomalous and 
non-anomalous spontaneously broken $U(1)$ symmetries. The hidden sector is assumed 
to consist of the universal dilaton, K\"ahler class and complex structure moduli, which are 
supposed to break supersymmetry, and a minimal set of Higgs fields which compensate 
the Fayet--Iliopoulos terms. We leave the superpotential that is supposed to stabilize 
the hidden sector fields unspecified, but we carefully take into account the relations 
implied by gauge invariance and the constraints required for the existence of a metastable 
vacuum with vanishing cosmological constant. The results are parametrized in terms of a 
constrained Goldstino direction, suitably defined effective modular weights, and the $U(1)$ 
charges and shifts. We show that the effect induced by vector multiplets strongly depends on the 
functional form of the K\"ahler potential for the Higgs fields. We find in particular that whenever 
these are charged matter fields, like in heterotic models, the effect is non-trivial, whereas 
when they are shifting moduli fields, like in certain intersecting brane models, the effect 
may vanish.

\vspace{5pt}
\end{quote}

\renewcommand{\theequation}{\thesection.\arabic{equation}}

\newpage

\section{Introduction}
\setcounter{equation}{0}

Superstring models represent a very appealing possibility for the microscopic framework 
underlying supersymmetric extensions of the standard model. In this respect, a crucial 
question concerns the way in which spontaneous supersymmetry breaking is realized. 
The standard paradigm is that this breaking occurs in a hidden sector and is then transmitted
only through suppressed gravitational interactions to the visible sector containing the 
supersymmetric extension of the standard model \cite{gravmed1,gravmed2,gravmed3}
(see also \cite{gravun1,gravun2,gravun3}). Since the scale of supersymmetry 
breaking must be much lower than the Planck scale, it is possible to study the problem
within a low-energy effective supergravity description. The structure of the soft supersymmetry 
breaking terms depends however on the structure of certain higher-dimensional operators, 
as well as on the direction of supersymmetry breaking. A particularly important issue is the 
flavor structure of the soft scalar masses. For phenomenological reasons, these should be nearly 
universal, or suitably aligned with the ordinary fermion masses. It is then natural to explore how 
this could come about.

A natural candidate for the hidden sector of string models is that of the universal neutral moduli 
fields, whose scalar Vacuum Expectation Values (VEV) control the strength of the coupling and 
the geometry of the internal space \cite{KL,BIM,BIMS}. An interesting point about this assumption 
is that the K\"ahler potentials for these moduli fields are determined by dimensional reduction and 
are universal, at leading order in the weak-coupling and low-energy expansions. They turn out to 
define constant curvature coset K\"ahler manifolds of the type $G/H$, where $G$ is a group of global 
isometries and $H$ is a local stability subgroup. Moreover, also the couplings between matter fields 
and moduli fields, which control soft scalar masses, have in general a special structure. 
More precisely, the leading higher-dimensional operators that are relevant for these soft 
masses can be parametrized in terms of certain effective modular weights for matter fields,
which are constant in the simplest cases but can in general depend on the moduli fields. 
The superpotential that could be at the origin of the spontaneous breaking of supersymmetry 
is on the other hand less understood. One can however leave it unspecified, and only assume 
that it produces a scalar potential admitting a stationary point where the energy is approximately 
zero and all the moduli are stabilized with a positive squared mass. The situation can then be 
parametrized in terms of the Goldstino vector of auxiliary fields, which has a length that is 
fixed by the condition of vanishing cosmological constant and a direction that can a priori 
be arbitrary but will be constrained to certain cones by the metastability condition 
\cite{grs1,grs2,grs3}. The structure of soft terms can then be parametrized in terms of the 
Goldstino direction and the effective modular weights. 

An important additional ingredient, which occurs in essentially all known models and can 
significantly change the situation, is the presence of additional $U(1)$ gauge symmetries. 
For phenomenological reasons, these must be spontaneously broken, with a sufficiently 
large mass for the corresponding gauge bosons. A beautiful way to make this breaking 
natural is provided by anomalies in these extra $U(1)$ symmetries, which are also pretty 
endemic. These anomalies are cancelled through the Green--Schwarz mechanism \cite{GS}, 
which has a slightly different form in different models, but rests on the same general structure. 
The basic point is that one-loop effects do not only induce anomalies, but also corrections to 
the K\"ahler potential, which force some of the moduli $M^i$, which were neutral at the tree level, 
to acquire at one-loop order a non-trivial behavior under the $U(1)$ transformations, shifting 
them by some constants $\delta_{ia}$. This implies then a non-trivial gauge variation of the 
gauge-kinetic functions depending on these moduli, which cancels the anomalies. 
This mechanism implies the emergence of moduli-dependent  Fayet--Iliopoulos terms, which 
act as a sources for the $D$-term contribution to the scalar potential once these moduli 
are stabilized at constant values \cite{4DGS}. In all known models, there exist however additional 
chiral multiplets with appropriate $U(1)$ charges that get VEV to approximately compensate such 
sources in the $D$-term potential, in such a way that the main effect of the Fayet--Iliopoulos terms is 
to break the $U(1)$ symmetries and to preserve supersymmetry, rather than vice-versa. In such a situation, 
the anomalous $U(1)$ symmetries are necessarily non-linearly realized, and the corresponding 
gauge bosons are massive. The Higgs fields are mostly a mixture of the charged fields taking 
VEV, with a small contamination from the moduli that can be neglected at leading order in 
$\delta_{ia}$, and the squared mass of the $U(1)$ vector bosons is of the order of a loop factor 
times the Planck mass squared. Furthermore, also non-anomalous $U(1)$ factors can be 
forced to be broken, if the mentioned Higgs fields have non-vanishing charges with respect to them.
In fact, in minimal situations where there are as many Higgs fields as $U(1)$ symmetries, the 
VEV of the Higgs fields are entirely determined by the various $D$-flatness conditions, which 
do or do not have moduli-dependent Fayet--Iliopoulos source terms depending on whether the 
corresponding symmetry is anomalous or not.

It has been appreciated for already some time that the presence of vector multiplets with masses 
below the Planck scale induces important additional effects mixing the visible and the hidden 
sectors, if both the visible and the hidden sector fields have non-vanishing charges 
\cite{Drees,HagelinKelley}. In fact, they give a sizable additional contribution to the soft scalar 
masses $m^2$, in spite of the fact that the auxiliary fields of these vector multiplets are suppressed 
as $D \sim m_{3/2}^2 M_{\rm P}^2/M_V^2$, and thus much smaller than the auxiliary fields 
of the hidden sector chiral multiplets, which are of order $F \sim m_{3/2} M_{\rm P}$. The reason is 
that if both the visible and the hidden sector fields are charged, there is a contribution to $m^2$ 
that is proportional to $D$, coming from the direct minimal renormalizable coupling to the vector 
multiplet, and which is of the same size as the ordinary contribution proportional to $F^2/M_{\rm P}^2$, 
coming from non-renormalizable effective interactions with the hidden sector chiral multiplets induced 
by gravity. This additional contribution can also be understood as coming from additional operators 
that are induced in the low-energy effective theory for the light chiral multiplets when integrating out 
the heavy vector multiplets, with $D \sim F^2/M_V^2$. In general this effect cannot be neglected, 
and brings a dependence on the $U(1)$ charges in the soft scalar masses. This represents an 
additional potential source of flavor non-universality, since the charges are not necessarily flavor 
universal, and the broken symmetries might even be related to flavor physics. It is therefore 
important to study the structure of soft scalar masses in the presence of such non-linearly 
realized $U(1)$ symmetries, to determined how much these change the situation with respect 
to the flavor problem. 

The effects on soft terms of a heavy vector field associated to an anomalous $U(1)$ symmetry 
were first explored in \cite{BinetruyDudas, DvaliPomarol, Barreiroetall}, mostly within a simplified
effective set up and assuming that supersymmetry breaking originates from gaugino condensation. 
The structure of soft scalar masses in string models with anomalous $U(1)$ factors, with a more 
general supersymmetry breaking sector and no assumption on the origin of supersymmetry 
breaking, has instead been first studied in \cite{Dudas1,Dudas2} and in \cite{KawamuraD1, 
KawamuraD2,KawamuraD3}, with slightly different points of view. However, in these papers 
it was implicitly assumed that the hidden sector superpotential does not depend on 
the charged fields Higgsing the $U(1)$ symmetries. We believe that this is not appropriate, 
because the moduli fields participating in the Green--Schwarz mechanism shift under gauge 
transformations, and a non-trivial superpotential leading to generic $F$ auxiliary fields can be 
gauge-invariant only if it depends on holomorphic gauge-invariant combinations of moduli fields 
and charged Higgs fields. In fact, in the minimal case, the relative effect of the Higgs fields 
compared to the moduli fields in supersymmetry breaking is completely fixed by gauge invariance. 
Subsequently, a proper computation of the form of soft scalar masses in the presence of an arbitrary 
but gauge-invariant superpotential has been performed in \cite{ADM}, for the minimal situation involving 
a modulus field and a matter Higgs field transforming respectively with a shift and a phase under an 
anomalous $U(1)$ symmetry. More recently, this analysis was generalized to include also additional 
non-shifting moduli \cite{ChoiDFF}. But a general and complete discussion of the detailed form of the 
scalar masses in string models with possibly several broken $U(1)$ symmetries and its implications 
for the flavor problem is still missing. 

The aim of this paper is to examine the structure of soft scalar masses in various kinds of string 
models involving spontaneously broken $U(1)$ symmetries, leaving the superpotential that is supposed 
to stabilize the moduli unspecified, but paying attention to the constraints implied by its gauge 
invariance. In our analysis, we will take into account the constraints that are put on the Goldstino 
direction by the requirement that all the moduli fields should be stabilized with a positive squared mass 
\cite{grs1,grs2,grs3}. We will also generalize previous studies to cases involving several anomalous 
and non-anomalous $U(1)$ symmetries, and Higgs fields that are either matter fields with a canonical 
quadratic K\"ahler potential or moduli fields with a non-canonical K\"ahler potential. Finally, we will 
examine more closely the typical situations arising in heterotic and brane models. In heterotic orbifold 
models, the stabilization of the dilaton modulus implies a Fayet--Iliopoulos term that is compensated by 
a canonical charged matter Higgs fields. This leads unavoidably to a non-trivial $D$-term contribution to soft 
scalar masses. In intersecting brane models, on the other hand, the stabilization of the dilaton and 
complex structure moduli may or may not generate Fayet--Iliopoulos terms, depending on whether the 
angles preserve or not some $U(1)$ R-symmetry. In the second case, the net Fayet--Iliopoulos terms induced 
by the moduli fields are again compensated by canonical charged matter Higgs fields, and the situation is 
very similar to that of heterotic models. In the former case, on the other hand, one can interpret the complex 
structure moduli as non-canonical shifting moduli Higgs fields compensating the Fayet--Iliopoulos terms 
induced by the dilaton, and the situation is radically different. Most importantly, there exist in this case 
situations where the $D$-term contribution to soft scalar masses vanishes identically, as a consequence 
of the related functional forms of the K\"ahler potentials for the various involved moduli fields.

The paper is organized as follows. In section 2, we review some general results concerning scalar
masses in supergravity models. In section 3, we examine more specifically the effects of heavy vector 
multiplets. In sections 4 and 5 we study the $F$- and $D$-term contributions to soft scalar masses in models 
with $U(1)$ symmetries broken at a high scale, for general situations where the Fayet--Iliopoulos terms 
are approximately cancelled by the VEV of a minimal set of Higgs fields with respectively a canonical
and non-canonical K\"ahler potential. In section 6 we apply these general results to string models
with a supersymmetry breaking sector identified with the untwisted moduli sector, focusing in particular
on heterotic orbifolds and intersecting brane models. Finally, in section 7 we summarize our conclusions.

\section{Soft scalar masses in supergravity models}
\setcounter{equation}{0}

We will start by briefly reviewing some of the salient features of supergravity models that will be needed 
for our analysis \cite{sugra1,sugra2,sugra3,sugra4,sugra5}. Let us consider more specifically a generic 
supergravity theory involving some chiral multiplets $Z^r$ and some $U(1)$ vector multiplets $V^a$. 
Setting $M_{\rm P} = 1$, the theory is specified by a real K\"ahler and gauge-invariant function 
$G = K + {\rm ln} |W|^2$, depending on $Z^r$, $\bar Z^r$ and $V^a$, and a holomorphic gauge-invariant 
gauge-kinetic function $f_{ab}$, depending on $Z^r$. Derivatives with respect to the fields $Z^r$, $\bar Z^r$ 
and $V^a$ are denoted by lower indices $r$, ${\bar r}$ and $a$, which are raised through the inverse of the 
K\"ahler metric $g_{r \bar s} = G_{r \bar s}$, and the inverse of the real part of the gauge kinetic function 
$h_{ab} = {\rm Re}\,f_{ab}$. 

The $U(1)$ gauge transformations are specified by holomorphic Killing vectors $X_a$, 
generating isometries of the scalar manifold, with components $X_a^r$ depending on 
the fields $Z^r$. More precisely, the supergauge transformations of the chiral and vector 
superfields are given by $\delta Z^r = \Lambda^a X_a^r$ and 
$\delta V^a = - i \,(\Lambda_a  - \bar \Lambda_a)$. The function $G$ must be invariant. 
This implies that $G_a = - i \, X_a^r G_r = i \, X_a^{\bar r} G_{\bar r}$. Taking derivatives, 
one then also deduces that $X_{a r} = - i \, G_{a r}$, showing that $G_a$ represent
Killing potentials. The function $f_{ab}$ must instead have a gauge variation that 
matches possible residual quantum anomalies $Q_{abc}$: $\delta f_{ab} = i\, \Lambda^c Q_{abc}$. 
This implies that $h_{a b r} = i/2 \, X^c_r \, Q_{abc}$.

The potential for the complex scalar fields of the chiral multiplets, which is the crucial quantity 
controlling spontaneous supersymmetry breaking, has the following form:
\bea
\a\a V = e^G \Big(G^r G_r - 3 \Big) + \frac 12 G^a G_a \,.
\eea
The flatness condition of vanishing cosmological constant is that $V = 0$ on the vacuum, and 
it implies:
\bea
\a\a - 3 + G^r G_r + \frac 12 \, e^{-G} G^a G_a = 0\,.
\label{flatness}
\eea
The stationarity conditions correspond to requiring that $\nabla_r V = 0$, and read:
\bea
\a\a G_r + G^s \nabla_r G_s + e^{-G} \Big[G_{ar} G^a
+ \frac 12\, \Big(h_{abr} - h_{ab} G_r \Big)G^a G^b \Big] = 0\,.
\label{stationarity}
\eea

The Hermitian block of the mass matrix for small fluctuations of the scalar fields around the 
vacuum can be computed as $m_{r \bar s}^2 = \nabla_r \nabla_{\bar s} V$, and is found to be 
\cite{FKZ,dudasvempati,grs3}:
\bea
\a\a m_{r \bar s}^2 = e^G \Big[g_{r \bar s} + \nabla_r G_t \nabla_{\bar s} G^t \!
- R_{r \bar s t \bar u} \, G^t G^{\bar u} \Big] \nn \\[1mm]
\a\a \hspace{31pt} \,+\, \Big[h^{ab} G_{a r} G_{b \bar s} 
+ \Big(G_{a r \bar s} - 2\, h^{bc} h_{ab(r} G_{c \bar s)}  - 2\, G_{a (r} G_{\bar s)}\Big) G^a \label{mijbar} \\[0mm]
\a\a \hspace{48pt} - \frac 12\, \Big(h_{ab}\, g_{r \bar s} - 2\,h^{cd} h_{a c r} h_{b d \bar s} 
- 2\, h_{ab(r} G_{\bar s)} - h_{ab} \, G_r G_{\bar s}  \Big)\, G^a G^b \Big] \,. \nn
\eea
The symmetric mass matrix for the vector fields has the form:
\bea
\a\a M_{ab}^2 = 2\, g^{r \bar s} G_{a r} G_{b \bar s} \,.
\label{Mab}
\eea
Finally, the gravitino mass is given by $m_{3/2} = e^{G/2}$.

The auxiliary fields controlling supersymmetry breaking are determined by their equations 
of motion and are given by 
\bea
\a\a F_r = - e^{G/2} \,G_r \,,\quad\; D_a = - G_a \,.
\eea 
These fields are however not independent from each other. A first kinematical relation 
between them, which holds at any point of field space and represents a constraint
due to gravity, is implied by the gauge invariance of the function $G$. It reads: 
\bea
\a\a D_a = -\,i \frac {X_a^r}{m_{3/2}} F_r \,. 
\label{kin}
\eea
A second dynamical relation between them, which holds only at stationary points 
of $V$ and exists independently of gravity, comes from the stationarity conditions along 
the complex partners of the would-be Goldstone directions. Indeed, contracting 
the stationarity conditions with $X_a^r$ and taking the imaginary part, one deduces 
that \cite{KawamuraDFF} (see also \cite{ChoiDFF,grs3}):
\bea
\a\a G_{a r \bar s} \, F^r F^{\bar s}  
+ \frac 12\,\Big[ M^2_{ab} + 2 \Big(F^r F_r - m_{3/2}^2 \Big) h_{ab} \Big]\, D^b
-\, \frac 12 \, Q_{abc} D^b D^c = 0 \,.
\label{dyn}
\eea

Let us now subdivide the chiral multiplets $Z^r$ into visible sector multiplets $Q^\alpha$
and hidden sector multiplets $X^i$. The visible sector multiplets $Q^\alpha$ are distinguished 
by the fact that all their components have vanishing VEV. This implies in particular that 
$G_\alpha = 0$, $g_{\alpha \bari} = 0$ and $\nabla_\alpha G_i = \nabla_i G_\alpha = 0$
on the vacuum. We will furthermore focus on matter fields in chiral representations of the 
visible gauge group that do not admit holomorphic quadratic invariants. This implies that 
$\nabla_\alpha G_\beta = 0$. We will also require that extra gauge symmetries are at 
most linearly realized on them, so that $G_{a \alpha} = 0$ on the vacuum. Finally, since 
the gauge kinetic function must be invariant under the visible gauge group, one also has 
$h_{a b \alpha} = 0$. For the hidden sector multiplets $X^i$, on the other hand, 
we do not impose any particular constraint for the moment.

The hidden sector fields can have a pretty generic dynamics. There are however two 
strong requirements that must be imposed in order to get a satisfactory situation at a 
certain stationary point. The first is the flatness condition of vanishing vacuum energy,
and implies a restriction on the length of the Goldstino vector:
\bea
\a\a g_{i \barj} \, F^i F^{\barj} = 3\, m^2_{3/2} - \frac 12 D^a D_a \,.
\eea
The second is the stability condition of positivity of the mass matrix for small fluctuations 
around the vacuum. As shown in \cite{grs1,grs2,grs3}, a necessary condition for this to 
happen is that the Hermitian block $m^2_{i \barj}$ be positive along the direction $G^i$,
implying the constraint $m^2_{i \barj} \, G^i G^{\barj} > 0$. This leads to the following condition,
which restricts the direction of the Goldstino vector \cite{grs3}:
\bea
\a\a R_{i \barj p \bar q} \, F^i F^{\barj} F^p F^{\bar q} \le 6 \, m^4_{3/2}
+ \Big(M^2_{ab} - 2 \,m^2_{3/2}\,h_{ab} + h^{cd} h_{a c i} h_{b d \barj} \, F^i F^{\barj} \Big)\, D^a D^b 
\label{stability} \\
\a\;\a \hspace{90pt} +\,\frac 34\,m_{3/2} \, Q_{abc} \, D^a D^b D^c 
- \frac 12 \Big(h_{ab} h_{cd} - \frac 12\, h_{ab}^{\;\;\;k} h_{cdk} \Big) D^a D^b D^c D^d \,.\nn
\eea

The additional vector fields have a mass matrix which is automatically positive definite and
by assumption entirely generated by the hidden sector fields. It takes the form:
\bea
\a\a M_{ab}^2 = 2\, g^{i \barj} G_{a i} G_{b \barj} \,.
\eea

The visible sector fields get soft masses through higher-dimensional operators mixing 
them to the hidden sector fields. Under the assumptions made above, the non-trivial 
Hermitian block of the scalar mass matrix takes the following form:
\bea
\a\a m_{\alpha \bar \beta}^2 = 
- \Big(R_{\alpha \bar \beta p \bar q} - \frac 13\, g_{\alpha \bar \beta}\, g_{p \bar q} \Big) F^p F^{\bar q}
- \Big(G_{a \alpha \bar \beta} - \frac 13\, g_{\alpha \bar \beta} \, G_a \Big) D^a \,.
\label{mabbar}
\eea
One can rewrite this expression in an alternative form, which uses the parametrization 
$G = - 3\, {\rm ln} \, (- \Omega /3)$ that is naturally suggested by the superconformal 
approach. With the same assumptions as before, one can easily verify that at the vacuum 
the following relations hold true:
\bea
\a\a R_{\alpha \bar \beta p \bar q} - \frac 13\, g_{\alpha \bar \beta}\, g_{p \bar q} = 
- \frac 3\Omega \, \Big(\Omega_{\alpha \bar \beta p \bar q}
- \Omega^{-1 \gamma \bar \delta}\, \Omega_{\alpha p \bar \delta}\, \Omega_{\bar \beta \bar q \gamma} \Big) \,, \\
\a\a G_{a \alpha \bar \beta} - \frac 13\, g_{\alpha \bar \beta} \, G_a = 
- \frac 3\Omega \, \Omega_{a \alpha \bar \beta} \,.
\eea
Plugging these expressions into eq.~(\ref{mabbar}) one finds:
\bea
\a\a m_{\alpha \bar \beta}^2 = \frac 3\Omega \bigg[\Big(\Omega_{\alpha \bar \beta p \bar q}
- \Omega^{-1 \gamma \bar \delta}\, \Omega_{\alpha p \bar \delta}\, \Omega_{\bar \beta \bar q \gamma} \Big) 
F^p F^{\bar q} + \Omega_{a \alpha \bar \beta}\, D^a \bigg]\,.
\label{mabbarbis}
\eea
This expression can also be obtained directly from the superconformal formulation. 
The auxiliary field $F$ of the compensator chiral multiplet does not give any contribution
\footnote{Note that this is a consequence of the vanishing of the cosmological constant.
In the presence of a cosmological constant $V$, one would find a contribution 
$- 2\, \Omega_{\alpha \bar \beta}/ \Omega \, V$ or equivalently $2/3 \, g_{\alpha \bar \beta}\, V$, 
which is entirely due to the Weyl rescaling of the potential by the factor $|\phi|^4$, which for the 
Einstein frame is equal to $e^{2 G / 3}$.}. 
The auxiliary fields $F^i$ of the hidden sector chiral multiplets give the first term directly 
and the second through their mixing with the auxiliary fields $F^\alpha$ of the visible sector
chiral multiplets. The auxiliary fields $D^a$ of the vector multiplets give the last term. Finally, 
the overall factor is due to the Weyl rescaling associated to the gauge choice $\phi = e^{G/6}$.

The two equivalent expressions (\ref{mabbar}) and (\ref{mabbarbis}) for the soft masses are
both useful, in different respects. The form (\ref{mabbar}) exhibits an interpretation of the masses
in terms of K\"ahler geometry. It shows for instance that the $F$-term contribution vanishes for 
maximally symmetric spaces of constant curvature scale equal to $2/3$, corresponding to "no-scale" 
models \cite{noscale}, for which $R_{\alpha \bar \beta p \bar q} = 1/3 \, g_{\alpha \bar \beta} \, g_{p \bar q}$, 
and similary that the $D$-term contribution vanishes whenever $G_{a i \barj} = 1/3 \, g_{i \barj} \,G_a$. 
The form (\ref{mabbar}) gives instead an interpretation of the masses that is more physically connected 
to the rigid limit intuition. It shows that the $F$-term contribution vanishes whenever the function $\Omega$ 
is separable into two distinct parts describing the visible and the hidden sectors, corresponding to the 
so-called sequestered situation \cite{sequestered}, and similarly that the $D$-term contribution vanishes 
whenever the function $\Omega$ does not contain any minimal coupling between the visible sector fields 
and the vector fields.

\section{Effects of heavy vector multiplets}
\setcounter{equation}{0}

The effect of vector multiplets relative to that of chiral multiplets substantially simplifies whenever
the mass eigenvalues of the vector fields are all much larger than the gravitino mass: $M_a \gg m_{3/2}$. 
We also assume that $M_a \lsim 1$, since the vectors are kept in the effective theory below the Planck scale,
and that $m_{3/2} \ll 1$, in order for supersymmetry to help explaining the hierarchy between the Fermi 
and the Planck scale. The above conditions for the vector masses are usually verified for the vector fields 
associated to anomalous $U(1)$ symmetries with a Green--Schwarz mechanism in string models. 
We will therefore focus on this situation of heavy vector fields from now on, and proceed to derive some 
simple general results emerging in this limit. Most importantly, it turns out that the vector auxiliary fields 
$D^a$ become small compared to the chiral auxiliary fields $F^i$, and the vector multiplet dynamics 
decouples from the supersymmetry breaking dynamics, which is dominated by chiral multiplets, leaving 
only simple corrections. These corrections are formally sub-leading, but they can nevertheless be relevant.

In the limit $M_a \gg m_{3/2}$, the flatness condition (\ref{flatness}) and the dynamical 
relation (\ref{dyn}) imply together that $F^i \sim {\cal O}(m_{3/2})$ and 
$D^a \sim {\cal O}(m_{3/2}^2/M_a^2)$, with:
\bea
\a\a D^a \simeq - 2\, M^{-2ab} G_{b i \barj} \, F^i F^{\barj} \,.
\label{dynnew}
\eea
The kinematical relation (\ref{kin}) implies then that the VEV of the scalar fields 
arrange in such a way that their contributions to $D^a$ cancel each other at 
the leading order ${\cal O}(m_{3/2})$, and leave only a subleading effect 
of order ${\cal O}(m_{3/2}^2/M_a^2)$ as implied by (\ref{dynnew}). In other words, 
we get following the approximate $D$-flatness conditions:
\bea
\a\a X_a^i F_i \simeq 0 \,.
\label{kinnew}
\eea
In this situation, the flatness and stability constraints on the Goldstino vector of 
auxiliary fields can be approximated as:
\bea
\a\a F^iF_i \simeq 3\, m_{3/2}^2 \,,
\label{flatnessnew}
\eea
and 
\bea
\a\a R_{i \barj p \bar q} \, F^i F^{\barj} F^p F^{\bar q} - M^2_{ab} \, D^a D^b \lsim 6 \, m^4_{3/2} 
\label{stabilitynew} \,.
\eea
The soft masses can in turn be rewritten in the form:
\bea
\a\a m_{\alpha \bar \beta}^2 \simeq g_{\alpha \bar \beta} \, m^2_{3/2}
- R_{\alpha \bar \beta p \bar q}  F^p F^{\bar q} - G_{a \alpha \bar \beta} D^a \,.
\label{mabbarnew}
\eea

Whenever the vector fields are much heavier that the gravitino, as in the above situation,
one can actually account for their leading-order effect in a much simpler way. Indeed,
the scale of gauge symmetry breaking is then much higher than the scale of supersymmetry 
breaking. One can then first integrate out the heavy vector multiplets to define a simpler 
effective theory with only chiral multiplets, which can be used to describe supersymmetry 
breaking. The vector multiplets can be integrated out directly at the level of superfields. 
In doing so, one can neglect terms with supercovariant derivatives, and freeze the 
superconformal compensator to $\Phi = e^{G/6}$. The relevant equation of motion of the 
vector superfield in terms of the chiral superfield is then simply that $G_a \simeq 0$. One 
must then choose a supersymmetric gauge. It is convenient to impose for this the gauge-fixing 
condition $G_a \simeq 0$, where the dependence on the vector superfields is now discarded, 
which corresponds to an approximate unitary gauge. This allows to consistently assume that 
the superfields $V^a$ are small and to make an expansion of the action in powers of $V^a$, to 
linearize the problem. At leading order in $V^a$, the equations of motion yields 
then \cite{grs3}:
\bea
\a\a V^a \simeq - 2\, M^{-2ab} G_b 
\label{Vsol} \,.
\eea
Taking the $D$ component of this expression, where $M^{2}_{ab}$ is treated as a number 
and $G_b$ is taken to depend only the chiral superfields, one recovers then the dynamical relation 
(\ref{dynnew}). One can also plug back the solution (\ref{Vsol}) into the quadratic approximation 
to the $V^a$-dependent terms in $G$, namely $\Delta G = G_a V^a + M^2_{ab}/4 \,V^a V^b$,
to deduce the form of the leading order correction that is induced by the heavy vector fields 
on the dynamics of the light chiral multiplets:
\bea
\a\a \Delta G \simeq - M^{-2ab} G_a G_b \,. 
\eea
It is straightforward to compute what is the corresponding correction to the K\"ahler curvature 
tensor. One finds:
\bea
\a\a \Delta R_{i \barj p \bar q} \simeq 
- 2 \,M^{-2ab} \Big(G_{a i \barj} \,G_{b p \bar q} + G_{a i \bar q} \,G_{b p \barj}\Big) \,, \\
\a\a \Delta R_{\alpha \bar \beta i \barj} \simeq 
- 2 \,M^{-2ab} G_{a \alpha \bar \beta} \,G_{b i \barj}  \,. 
\eea
Using these expressions, as well as the relation (\ref{dynnew}), we can then interpret the leading 
corrections depending on $D^a$ in eqs.~(\ref{stabilitynew}) and (\ref{mabbarnew}) as emerging 
from the corrections to the curvature tensors in the terms depending on $F^i$.

It is now important to observe that the corrections induced by the heavy vector multiplets in 
eqs.~(\ref{stabilitynew}) and (\ref{mabbarnew}) are in principle both significant. In fact, 
we will now see in somewhat more detail that in string models, where the squared masses
$M^2_{ab}$ are induced by Higgs fields and the symmetry breaking scale is around the 
Planck scale, that is $M^2_{ab} \sim {\cal O}(1)$, all the terms in eqs.~(\ref{stabilitynew}) 
and (\ref{mabbarnew}) are of the same order of magnitude, as far as the scaling with 
dimensionfull quantities is concerned. The only particularity of the terms induced 
through the vector multiplets is that they are proportional to the charges of the involved 
fields. These can happen to be small, if they reflect a non-minimal coupling induced at 
a subleading order in perturbation theory.

\section{Models with canonical Higgs fields}
\setcounter{equation}{0}

Let us consider first the case of $U(1)$ symmetries that are realized through rephasings on the 
matter fields $Q^\alpha$ and some gauge symmetry breaking fields $H^x$ with charges 
$q_{\alpha a}$ and $q_{x a}$, and through shifts $\delta_{i a}$ on the supersymmetry breaking 
fields $M^i$. This means that the Killing vectors have components that are given by 
$X_a^\alpha = i \, q_{\alpha a} Q^\alpha$, $X_a^x = i\, q_{x a} H^x$ and $X_a^i = -\,i\, \delta_{i a}$. 
For simplicity, we shall assume that $\delta_{i a} \ll 1$, as it turns out to be in most of the known 
string models, and evaluate all the formulae at leading order in $\delta_{i a}$. In such a situation, 
the would-be goldstone bosons that are absorbed by the gauge vectors are essentially linear 
combinations of the fields $H^x$, with only a small admixture of the fields $M^i$. We will consider 
the minimal situation where the number of gauge symmetry breaking fields $H^x$ with non-zero 
charges $q_{x a}$ is equal to the number of $U(1)$ vector fields $V^a$. The charge matrix $q_{x a}$ 
is then a square matrix that can be inverted, the inverse charge matrix $q_{a x}^{-1}$ 
being defined to satisfy the identities \footnote{In this section, we shall make sums 
over indices of type $a$ and $x$ explicit, to avoid confusions.}
$\sum_a q_{x a}^{-1} q_{a y} = \delta_{xy}$ and $\sum_x q_{a x} q_{x b}^{-1}  = \delta_{ab}$. 
We emphasize here that the shifts $\delta_{ia}$ can be pretty arbitrary. The only restriction 
that we shall impose is that they are not all zero, the non-vanishing ones setting the 
gauge symmetry breaking scales. This means that there can be an arbitrary 
but non-zero number of anomalous $U(1)$ symmetries and an arbitrary number of 
non-anomalous  $U(1)$ symmetries.

For concreteness, let us take as starting point the following general form of 
the K\"ahler potential defining the effective supergravity Lagrangian, with a canonical 
term for the Higgs fields:
\bea
\a\a K = \hat K \Big(M^i + \bar M^i - \sum_a \delta_{i a} V^a\Big)
+ \sum_{x} \tilde K_{x \bar x} \Big(M^i + \bar M^i - \sum_a \delta_{i a} V^a\Big) \, 
\bar H^x e^{\sum_a \! q_{x a} V^a} \! H^x \nn \\
\a\;\a \hspace{26pt} + \sum_\alpha K_{\alpha \bar \alpha} \Big(M^i + \bar M^i - \sum_a \delta_{i a} V^a\Big) \, 
\bar Q^\alpha e^{\sum_a \! q_{\alpha a} V^a} \! Q^\alpha + \dots \,.
\label{Kcanonical}
\eea
The superpotential is instead left to be a generic gauge-invariant 
holomorphic function of the shifting moduli $M^i$ and the charged Higgs fields $H^x$:
\bea
\a\a W = \hat W \Big(M^i + \sum_{a,x} \delta_{ia} q^{-1}_{ax} \, {\rm ln} \, H^x\Big) + \dots \,.
\eea
The dots in the above expressions denote possible additional terms that are of higher order 
in the fields $Q^\alpha$, whose VEV are vanishing by assumption, and in the fields $H^x$, 
whose VEV will turn out to be small. 

Our aim is to compute the total contribution to soft scalar masses of the chiral multiplets of the 
visible sector, due to both the $F^i$, $F^x$ and the $D^a$ auxiliary fields of chiral and vector 
multiplets of the hidden sector. This computation was first done in \cite{KawamuraD1,KawamuraD2},
in the context of heterotic orbifold models with a single anomalous $U(1)$, and generalized to 
type I models with several $U(1)$'s in \cite{KawamuraD3}. 
In the following, we shall present a more general computation, which is valid for an arbitrary number 
of anomalous and non-anomalous $U(1)$'s, an arbitrary number of shifting and non-shifting moduli, 
and a generic supersymmetry breaking dynamics compatible with gauge invariance.
We will first do the computation along the lines of \cite{KawamuraD1,KawamuraD2,KawamuraD3},
but paying attention to the consistency constraints put by gauge invariance of the superpotential. 
We will then also show how the same result can be obtained in a simpler way by integrating out the 
heavy vector multiplets directly at the superfield level, along the lines of \cite{ADM,ChoiDFF}.

On the vacuum, and at leading order in $\delta_{ia}$, the non-trivial components of the K\"ahler metric 
in the visible and hidden sectors are given by:
\bea
\a\a g_{\alpha \bar \alpha} = K_{\alpha \bar \alpha} \,;\quad\; 
g_{i \barj} = \hat K_{i \barj} + \sum_x \tilde K_{x \bar x i \barj} \, |H^x|^2 \,,\quad\;
g_{x \bar x} = \tilde K_{x \bar x} \,,\quad\; 
g_{x \bari} = \tilde K_{x \bar x \bari} \, \bar H^x \,.
\label{metric}
\eea
Since the off-diagonal terms are small, it is possible to find simple expressions also 
for the various blocks of the inverse of the metric, at leading order in $H^x$. One finds 
\bea
\a\a g^{\alpha \bar \alpha} = g^{-1}_{\alpha \bar \alpha} \,;\quad\;
g^{i \barj} = g^{-1}_{i \barj} \,,\quad\;
g^{x \bar x} = g^{-1}_{x \bar x} \,,\quad\;
g^{x \bari} =  - \sum_k g^{-1}_{x \bar x} \, g_{k \bar x} \, g^{-1}_{k \bari} \,.
\label{ginverse}
\eea
The tensor quantities that enter the formula for the soft masses can be 
easily evaluated. At leading order in $\delta_{ia}$, one finds
$R_{\alpha \bar \alpha p \bar q} = g_{\alpha \bar \alpha}\, \partial_p \partial_{\bar q}\, 
{\rm ln}\, \big(K_{\alpha \bar \alpha} \big)$ and 
$G_{a \alpha \bar \alpha} = g_{\alpha \bar \alpha}\, q_{\alpha a}$.
The soft masses take therefore the form:
\bea
\a\a m_{\alpha \bar \alpha}^2 = - g_{\alpha \bar \alpha} 
\bigg[ \partial_p \partial_{\bar q} \, {\rm ln}\, \Big(K_{\alpha \bar \alpha} e^{- \hat K/3}\Big) F^p F^{\bar q} 
+ \sum_a q_{\alpha a} D^a \bigg] \,.
\label{softmass}
\eea
Our main task is now to compute the $D^a$'s in terms of the $F^i$'s, after exploiting the fact that 
in the minimal situation under consideration, gauge invariance completely fixes the VEV of the 
Higgs scalar fields $H^x$ and the corresponding auxiliary fields $F^x$ in terms of the VEV of 
the moduli fields $M^i$ and the corresponding auxiliary fields $F^i$.

\subsection{Relation between auxiliary fields}

The vector auxiliary fields can be computed through the relation $D_a = - G_a$ 
and take the following form:
\bea
\a\a D_a =  \sum_k \delta_{k a} \, \hat K_{k} - \sum_{x} q_{x a}\, g_{x \bar x} \, |H^x|^2 \,.
\label{formD}
\eea
The mass matrix $M_{ab}^2$ of the vectors fields is given by
\bea
\a\a M_{ab}^2 = 2 \sum_{i,j} g_{i \barj}\, \delta_{i a} \, \delta_{\barj b}
+ 2 \sum_{x} q_{x a} q_{x b} \, g_{x \bar x} \, |H^x|^2 \,.
\label{formM}
\eea
Assuming that the VEV of the fields $M^i$ are of order ${\cal O}(1)$, and recalling that 
the shifts $\delta_{ia}$ are of the order of a loop factor, and thus small but not tiny, we 
see from the first term in the above expression that one indeed naturally gets 
the situation $M_{ab}^2 \gg m_{3/2}^2$ described in section 3. 

At the stationary point of the potential, the VEV that the $D_a$'s actually develop 
depend quadratically on the auxiliary fields of all the non-trivially transforming chiral 
multiplets, and are given by the relation (\ref{dynnew}), which becomes now:
\bea
\a\a D^a = - 2 \sum_b M^{-2ab} \Big(G_{b i \barj} \, F^i F^{\barj} + G_{b x \bar x} \, F^x F^{\bar x} 
+ G_{b i \bar x} \, F^i F^{\bar x} + G_{b x \barj} \, F^x F^{\barj}  \Big) \,.
\label{expD}
\eea
Neglecting terms of order ${\cal O}(\delta_{i a} |H^x|^2)$, the coefficients appearing in this
expression are found to be
$G_{b i \barj} = \sum_{x} q_{x b} \tilde K_{x \bar x i \barj} |H^x|^2 \!-\! \sum_k \delta_{k b} \hat K_{k i \barj}$,
$G_{b x \bar x} = q_{x b} \tilde K_{x \bar x}$ and 
$G_{b x i} = q_{x b} \tilde K_{x \bar x i} \bar H^x$.
Substituting these expressions into eq.~(\ref{expD}), completing the squares and  
using also the expressions (\ref{metric}), one can rewrite the result as:
\bea
\a\a D^a = - 2\sum_b M^{-2ab} \bigg[
\Big(\sum_{x} q_{x b} \, |H^x|^2 \Big(\tilde K_{x \bar x i \barj} - \tilde K^{-1}_{x \bar x} 
\tilde K_{x \bar x i}  \tilde K_{x \bar x \barj} \Big)
-\, \sum_k \delta_{k b} \, \hat K_{i \barj k} \Big)\, F^i F^{\barj} \nn \\[-1mm]
\a\a \hspace{100pt}
+\, \sum_{x} q_{x b} \, g_{x \bar x} \, \tilde F^x \tilde F^{\bar x} \bigg] \,,
\label{evalDnew}
\eea
where 
\bea
\a\a \tilde F^x = F^x + g^{x \bar x} g_{\bar x i} F^i \,.
\eea

The VEV of the fields $H^x$ are related to the VEV of the fields $M^i$. Indeed, 
since the auxiliary fields $D_a$ must obtain a small VEV of ${\cal O}(m_{3/2}^2/M_{a}^2)$, 
the terms of ${\cal O}(m_{3/2})$ in the expression (\ref{formD}) must cancel out. 
This leading order approximate $D$-flatness condition $D_a \simeq 0$ implies
that $\sum_{x} q_{x a} \, g_{x \bar x} \, |H^x|^2 \simeq \sum_k \delta_{k a} \, \hat K_{k}$.
The VEV of the $H^x$'s are then completely determined in terms of the VEV of the $M^i$'s.
Indeed, multiplying this equation by $q_{y a}^{-1}$ and summing then over $a$, one extracts:
\bea
\a\a |H^x| \simeq g_{x \bar x}^{-1/2} v_x \,,
\label{gHH}
\eea
where
\bea
\a\a v_x =  \Big(\sum_{k,a} \delta_{k a} q_{a x}^{-1} \hat K_k\Big)^{1/2} \,.
\label{vx}
\eea
The mass of the vector fields (\ref{formM}) is dominated by the second term. 
The mass matrix and its inverse can then be approximately written as:
\bea
\a\a M_{ab}^2 \simeq 2 \sum_{x} q_{x a} q_{x b} v_x^2 \,,\quad\;
M^{-2ab} \simeq \frac 12 \sum_x q_{a x}^{-1} q_{b x}^{-1} v_x^{-2}\,.
\label{Mapprox}
\eea

The values of the auxiliary fields $F_x$ are similarly related to the values of the auxiliary 
fields $F_i$. Indeed, according to eq.~(\ref{kin}), the $D_a$'s can also be written as linear 
combinations of the $F_i$'s and $F_x$'s: $D_a = -\,i\, \big(X_a^x F_x + X_a^k F_k\big)/m_{3/2}$,
and the approximate $D$-flatness conditions $D_a \simeq 0$ imply thus also the 
approximate relations (\ref{kinnew}), which in our case read
$\sum_x q_{x a} H^x F_x \simeq \sum_k \delta_{k a} F_k$.
Multiplying this equation by $q_{y a}^{-1}$, summing over $a$ and using 
the relation (\ref{gHH}), one deduces then that the $F_x$ are completely determined 
in terms of the $F_i$:
\bea
\a\a F_x =  g_{x \bar x}^{1/2} v_x^{-1} \sum_{k,a} \delta_{ka} q_{ax}^{-1} F_k \,.
\label{Fx}
\eea
It is now straightforward to derive the values of the auxiliary fields with upper indices, by using 
the inverse metric (\ref{ginverse}). Since $F_x$ is suppressed by a factor of order ${\cal O}(H^x)$ 
with respect to $F_i$, it is enough to keep ${\cal O}(H^x)$ terms only in $F^x$, and neglect them in $F^i$. 
The non-vanishing off-diagonal component $g^{x \barj}$ is thus relevant only in the computation 
of $F^x$, and after using $g^{x \barj}= - g^{x \bar x} g^{i \barj} g_{i \bar x}$, one finds
$F^x = g^{x \bar x} \big(F_{\bar x} - g^{i \barj} g_{i \bar x} F_{\barj} \big)$ and 
$F^i = g^{i \barj} F_{\barj}$. This implies then the simple relation $\tilde F^x = g^{x \bar x} F_{\bar x}$.
Finally, using the relation (\ref{Fx}), we see that gauge invariance forces the $\tilde F^x$'s to be linear 
combinations of the $F^i$'s: 
\bea
\a\a \tilde F^x =  g_{x \bar x}^{-1/2} v_x^{-1} \sum_{k,a} \delta_{ka} q_{ax}^{-1} 
\hat K_{k \barj} F^{\barj} \,.
\label{Fxup}
\eea
The above results on the relation between the auxiliary fields of the Higgs fields and those of 
the supersymmetry breaking fields are most conveniently summarized through the relation 
between $F^x/H^x$ and $F^i$. This can be easily deduced from the already mentioned relation 
$F^x = g^{x \bar x} \big(F_{\bar x} - g^{i \barj} g_{i \bar x} F_{\barj} \big)$, after using 
eqs.~(\ref{gHH}) and (\ref{Fx}). The result takes the following simple form:
\bea
\frac {F^x}{H^x} = \partial_i \, {\rm ln} \Big(\tilde K^{-1}_{x \bar x}  \sum_{k,c} \delta_{k c} q_{c x}^{-1} \hat K_k \Big)
F^i \,.
\label{FxHx}
\eea

We can now come back to the problem of evaluating eq.~(\ref{evalDnew}). Using the above 
results (\ref{gHH}), (\ref{Mapprox}) and (\ref{Fxup}) for $|H^x|$, $M^{-2ab}$ and $\tilde F^x$, 
one obtains:
\bea
\a\a D^a = \sum_{x} q_{a x}^{-1}  \bigg[\!-\Big(\tilde K^{-1}_{x \bar x} \tilde K_{x \bar x i \barj} 
- \tilde K^{-2}_{x \bar x} \tilde K_{x \bar x i}  \tilde K_{x \bar x \barj} \Big) \nn \\[-2mm]
\a\a \hspace{70pt} +\, \Big(\sum_{k,b} \delta_{kb} q_{bx}^{-1} v_x^{-2} \hat K_{i \barj k} 
- \! \sum_{k,l,b,c} \! \delta_{kb} q_{bx}^{-1} \delta_{l c} q_{c x}^{-1} v_x^{-4}
\hat K_{k \barj} \, \hat K_{i \bar l} \Big) \bigg] 
F^i F^{\barj}  \,. \hspace{20pt}
\eea
Recalling the expression (\ref{vx}) for $v_x$, this result can finally be rewritten in the 
very simple form
\bea
\a\a D^a = \sum_x q_{a x}^{-1} \, \partial_i \partial_{\barj} \, {\rm ln} \, 
\Big(\tilde K^{-1}_{x \bar x}  \sum_{k,c} \delta_{k c} q_{c x}^{-1} \hat K_k \Big)
F^i F^{\barj}  \,.
\label{resDa}
\eea

The above result can in fact be derived in a much simpler way by integrating out the 
heavy vector fields at the superfield level, along the lines of \cite{PomarolDimopoulos,Rattazzi} 
and \cite{ADM,ChoiDFF}. In the limit where $M_{a} \gg m_{3/2}$, supercovariant derivatives can be 
neglected and the equations of motion are entirely controlled by the K\"ahler potential. To start with, it 
is convenient to define the following new gauge-invariant combinations of chiral and vector 
fields:
\bea
\a\a Q^{\prime \alpha} = Q^\alpha \exp \bigg\{\!-\! \sum_{x,a} q_{\alpha a} q_{a x}^{-1} \, {\rm ln} \, H^x \bigg\} \,, \nn \\
\a\a M^{\prime i} = M^i + \sum_{x, a} \delta_{i a} q_{a x}^{-1} \, {\rm ln} \, H^x  \,, \\
\a\a V^{\prime a} = V^a + \sum_x q_{a x}^{-1} \Big({\rm ln} \, H^x + {\rm ln} \,\bar H^x \Big) \,. \nn
\eea
These allow to rewrite the K\"ahler potential and the superpotential with one less type of fields, 
namely without any $H^\prime_x$. In fact, this procedure can actually be interpreted as a gauge-fixing 
corresponding to setting $H^\prime_x = 1$, with the choice $\Lambda^a = i \sum_x q_{a x}^{-1} \, {\rm ln} \, H^x$ 
for the superfield gauge parameter. The K\"ahler potential can then be rewritten as
\bea
\a\a K = \hat K \Big(M^{\prime i} + \bar M^{\prime i} - \sum_a \delta_{i a} V^{\prime a} \Big)
+ \sum_{x} \tilde K_{x \bar x} \Big(M^{\prime i} + \bar M^{\prime i} - \sum_a \delta_{i a} V^{\prime a}\Big) \, 
e^{\sum_a \! q_{x a}V^{\prime a}} \nn \\
\a\;\a \hspace{26pt} 
+\, \sum_\alpha K_{\alpha \bar \alpha} \Big(M^{\prime i} + \bar M^{\prime i} - \sum_a \delta_{i a} V^{\prime a}\Big) \, 
\bar Q^{\prime \alpha} e^{\sum_a \! q_{\alpha a}V^{\prime a}} \! Q^{\prime \alpha} + \dots \,,
\eea
and the superpotential as
\bea
W = \hat W \Big(M^{\prime i}\Big) + \dots \,.
\eea
The superfield equations of motion for $V^{\prime a}$ are now easily computed by taking a derivative
of $K$ with respect to $V^{\prime a}$. At leading order in $\delta_{i a}$, and ignoring the visible sector fields,
one finds: 
\bea
\a\a \sum_x q_{x a} \tilde K_{x \bar x} \Big(M^{\prime i} + \bar M^{\prime i} \Big) 
e^{\sum_b q_{x b} V^{\prime b}} =  \sum_k \delta_{k a} \hat K_k
\Big(M^{\prime i} + \bar M^{\prime i} \Big) \,.
\eea
Notice that we could neglect the dependence on $V^{\prime a}$ in all the brackets. This is 
a radical simplification, which is crucial to be able to algebraically solve for $V^{\prime a}$.
The solution for $V^{\prime a}$ is finally obtained by using the inverse charge matrix $q_{a x}^{-1}$.
One finds:
\bea
V^{\prime a} = \sum_{x} q_{a x}^{-1} \, {\rm ln} \Big(\tilde K_{x \bar x}^{-1} \sum_{k,c} \delta_{k c} q_{c x}^{-1} \, 
\hat K_k \Big) \,.
\label{solVa}
\eea
Having found the solution of the superfield equation of motion for $V^{\prime a}$, it is now trivial 
to deduce the value of its $D^{\prime a}$ auxiliary component in terms of the auxiliary fields $F^{\prime i}$. 
Taking the $D$-component of (\ref{solVa}), one reproduces indeed the result (\ref{resDa}):
\bea
D^{\prime a} = \sum_{x} q_{a x}^{-1} \, \partial_i \partial_{\barj} \, {\rm ln} 
\Big(\tilde K_{x \bar x}^{-1} \sum_{k,c} \delta_{k c} q_{c x}^{-1} \, \hat K_k \Big) F^{\prime i} F^{\prime \barj}\,.
\eea

\subsection{Structure of the soft masses}

The soft masses of the visible fields are given by the expressions (\ref{softmass}) and (\ref{resDa}), and 
have the following rather simple and compact structure:
\bea
\a\a m_{\alpha \bar \alpha}^2 = - g_{\alpha \bar \alpha} \, \partial_i \partial_{\bar j}
\bigg[{\rm ln}\, \Big(K_{\alpha \bar \alpha} e^{- \hat K/3} \Big)
+ \sum_{a,x} q_{\alpha a} q_{a x}^{-1}  \, 
{\rm ln} \, \Big(\tilde K_{x \bar x}^{-1} \sum_{k,c} \delta_{k c} q_{c x}^{-1} \hat K_k \Big)
\bigg] F^i F^{\barj}\,. \hspace{10pt}
\label{soft} 
\eea
The expression (\ref{soft}) represents the result we were aiming at. It shows that the 
results of \cite{ADM,ChoiDFF} generalize in a pretty simple way to more 
general situations involving several $U(1)$ symmetries, which can be either anomalous 
or non-anomalous, with equally many gauge symmetry breaking fields, and an arbitrary
set of moduli fields, which can either shift or not shift under gauge transformations. 

The physical soft masses can be obtained by rescaling the above expression by the 
metric $g_{\alpha \bar \alpha}$, which appears as a non-trivial wave-function factor 
in the kinetic term. The result can then be rewritten in the following suggestive form:
\bea
\a\a m_\alpha^2 = m_{F \alpha}^{2} - \sum_{a,x} q_{\alpha a} q_{a x}^{-1} \, \Big(m_{F x}^2 - m_{Ax}^2\Big) \,,
\eea
where 
\bea
\a\a m_{F \alpha}^2 = - \partial_i \partial_{\barj}\,  {\rm ln} \Big(K_{\alpha \bar \alpha} e^{- \hat K/3}\Big) F^i F^{\barj}\,, \nn \\
\a\a m_{F x}^2 = - \partial_i \partial_{\barj}\,  {\rm ln} \Big(\tilde K_{x \bar x} e^{- \hat K/3} \Big)F^i F^{\barj} \,, \\
\a\a m_{Ax}^2 = - \partial_i \partial_{\barj}\,  {\rm ln} \Big(\sum_c q^{-1}_{c x} \hat K_{c} e^{- \hat K/3} \Big) F^i F^{\barj} \nn\,.
\eea
The first term is the usual $F$-term contribution. The second term isolates a general part of the $D$-term 
contribution that has the same form as for non-anomalous $U(1)$ symmetries, and consists in a 
linear combination of the $F$-term masses of the Higgs fields suitably weighted by their charges. 
Finally, the third contributions encodes essentially the non-trivial effects due to the fact that some 
$U(1)$ symmetries are anomalous, with some supersymmetry breaking fields transforming non-trivially 
under them.

One can also compute the form of the flatness and stability constraints (\ref{flatnessnew}) and 
(\ref{stabilitynew}). The effects of chiral multiplets correspond to the unique term in (\ref{flatnessnew}) 
and the first term in (\ref{stabilitynew}). In the case treated here, one should in principle include 
the effects of both the $M^i$ and the $H^x$ fields. However, the terms involving $F^x$ are 
negligible with respect to the terms involving $F^i$, meaning that that the effect of the $H^x$ 
fields can be neglected. The effects of the fields $V^a$ have already been neglected in 
(\ref{flatnessnew}) but lead to the potentially sizable second term in (\ref{stabilitynew}). 
This term can be computed by using the result (\ref{resDa}), and can then be reexpressed 
as a correction to the effective curvature for the fields $M^i$. However, it also turns out to be 
negligible, essentially because the "charges" induced for the moduli fields are only of 
order ${\cal O}(\delta_{ia})$ and thus small. In the end, the flatness and stability constraints 
can be written in terms of the metric and the curvature of the moduli space simply as:
\bea
\a\a g_{i \barj} \, F^i F^{\barj} \simeq 3\, m_{3/2}^2 \,, \\[1mm]
\a\a R_{i \barj p \bar q} \, F^i F^{\barj} F^p F^{\bar q} \lsim 6\, m_{3/2}^4 \,.
\eea

The physical components of the Goldstino directions can be obtained by suitably rescaling 
the auxiliary fields by the square root of the metric, which appears in front of their kinetic terms.

Notice finally that the rescaling procedure needed to switch to canonical scalar and auxiliary fields 
corresponds geometrically to switching to normal coordinates around the vacuum point. It is then 
natural to rewrite the above results in this local frame, whose indices will be denoted with capital 
letters. This is done simply by defining $Q^\alphacap = e^\alphacap_\alpha Q^\alpha$, $M^I = e^I_{i} M^i$, 
$H^X = e^X_x H^x$, and $V^A = e^A_a V^a$, with the help of the vielbeins $e^\alphacap_\alpha$, 
$e^I_i$, $e^X_x$ and $e^A_a$. In this new basis the metrics are trivial: 
$g \raisebox{-3pt}{$\alphacap \bar \betacap$} = \delta \raisebox{-3pt}{$\alphacap \bar \betacap$}$, 
$g_{I \bar J} = \delta_{I \bar J}$, $g_{X \bar Y} = \delta_{X \bar Y}$ and $h_{AB} = \delta_{AB}$. 
The physical soft scalar masses are then given in terms of the rescaled auxiliary fields $F^I = e^I_i F^i$ by
\bea
\a\a m_{\alphacap \bar \alphacap}^2 = \bigg[\frac 13 \, \delta_{I \bar J}
- R_{\alphacap \bar \alphacap I \bar J} +\, \sum_{A,X} q_{\alphacap A} q_{A X}^{-1}  \, 
\Big(R_{X \bar X I \bar J} - S_{X \bar X I \bar J} \Big)
\bigg] F^I F^{\bar J}\,, \hspace{10pt}
\label{softcan} 
\eea
and the flatness and stability conditions by
\bea
\a\a \delta_{I \bar J} \, F^I \bar F^{\bar J} \simeq 3\, m_{3/2}^2 \,,\;\; \\
\a\a R_{I \bar J P \bar Q} \, F^I \bar F^{\bar J} F^P \bar F^{\bar Q} \lsim 6\, m_{3/2}^4 \,,
\eea
where $q_{\alphacap A} \equiv q_{\alpha a}$, $q_{X A} \equiv q_{x a}$ and
\bea
\a\a R_{I \bar J P \bar Q} = e_I^i e_{\bar J}^{\barj} e_P^p e_{\bar Q}^{\bar q} 
\Big(\hat K_{i \barj p \bar q} - \hat K^{r \bar s} \hat K_{i p \bar s} \hat K_{\barj \bar q r} \Big) \,, \nn \\
\a\a R_{\alphacap \bar \alphacap I \bar J} = e_I^i e_{\bar J}^{\bar j} \, 
\partial_i \partial_{\barj} \, {\rm ln} \, \Big(K_{\alpha \bar \alpha} \Big) \,,\;\;
R_{X \bar X I \bar J} = e_I^i e_{\bar J}^{\barj} \, 
\partial_i \partial_{\barj} \, {\rm ln} \, \Big(\tilde K_{x \bar x} \Big) \,, \label{curvflat} \\
\a\a S_{X \bar X I \bar J} = e_I^i e_{\bar J}^{\barj} \, \partial_i \partial_{\barj} \, {\rm ln} \,
\Big(\sum_{k,c} \delta_{k c} q_{c x}^{-1} \hat K_k \Big) \,. \nn 
\eea

\subsection{Basic moduli}

Let us now illustrate the general result derived above in the particular case where 
the supersymmetry breaking fields are a set of universal string moduli, each spanning 
a distinct one-dimensional K\"ahler manifold, which in the low-energy limit has the form 
of an $SU(1,1)/U(1)$ coset space. In general, all these moduli can shift under 
the $U(1)$ symmetries, with arbitrary but small shifts $\delta_{ia}$. The K\"ahler potential 
has then the following structure:
\bea
\a\a \hat K = - \sum_i \,c_i\,{\rm ln}\,\Big(M^i + \bar M^i - \sum_a \delta_{ia} V^a \Big) \,.
\eea
Applying eqs.~(\ref{curvflat}), one easily finds:
\bea
\a\a R_{I \bar J P \bar Q} = \frac 2{c_I} \, \delta_{I \bar J P \bar Q} \,, \label {Rconst} \\[1mm]
\a\a S_{X \bar X I \bar J} = \frac 2{c_I} \, \frac {t_{IX}}{\big(\sum_K t_{KX}\big)} \, \delta_{I \bar J} 
- \frac 1{\sqrt{c_I c_J}} \,  \frac {t_{IX} \, t_{JX}}{\big(\sum_K t_{KX}\big)^2} 
\label{Snonconst} \,,
\eea
in terms of the following quantities, proportional to the inverse of the VEV of the 
moduli fields through a particular linear combination of their shift parameters: 
\bea
t_{I X} = \sum_a \delta_{ia} q_{ax}^{-1} c_i \big(M^i + \bar M^i \big)^{-1} \,.
\label{tvar}
\eea
The quantities $R_{\alphacap \bar \alphacap I \bar J}$ and $R_{X \bar X I \bar J}$ represent 
the unspecified mixed components of the curvature corresponding to the direct mixing 
between matter and Higgs fields with moduli fields. These generically depend on the VEV 
of the moduli fields, but there exist also special situation where they are constant. For instance,
if $K_{\alpha \bar \alpha} = \prod_i \big(M^i + \bar M^i - \sum_a \delta_{ia} V^a\big)^{n_{\alpha i}}$ 
and $K_{x \bar x} = \prod_i \big(M^i + \bar M^i - \sum_a \delta_{ia} V^a \big)^{n_{xi}}$, with constant 
modular weights $n_{\alpha i}$ and $n_{x i}$, like in modular-invariant heterotic models, then one 
finds simply $R_{\alphacap \bar \alphacap I \bar J} = - n_{\alphacap I}\, \delta_{I \bar J}$ 
and $R_{X \bar X I \bar J} = - n_{X I} \, \delta_{I \bar J}$. On the other hand, the quantity 
$S_{X \bar X I \bar J}$ has a fixed dependence on the moduli fields, which is non-trivial 
as soon as several fields shift under $U(1)$ symmetries. In the particular case of models 
with a single anomalous $U(1)$ and a single shifting modulus $M^0$, one has however 
$t_{IX} = t_{0X}\, \delta_{0I}$ and one finds thus a constant result given by 
$S_{X \bar X I \bar J} = c_0^{-1}\, \delta_{0 I \bar J}$, independently 
of the values of the shift and the charges.

\section{Models with non-canonical Higgs fields}
\setcounter{equation}{0}

Let us consider next the case of $U(1)$ symmetries that are realized 
through rephasings on the matter fields $Q^\alpha$ and through shifts $\eta_{xa}$
and $\delta_{i a}$ on the gauge symmetry breaking fields $M^x$ and on the supersymmetry 
breaking fields $M^i$. This means that the components of the Killing vectors defining the action 
of the $U(1)$ symmetries are given by $X_a^\alpha = i \, q_{\alpha a} Q^\alpha$, 
$X_a^x = -\,i\, \eta_{x a}$ and $X_a^i = -\,i\, \delta_{i a}$. For simplicity, we shall again assume 
for the moment that $\delta_{i a} \ll 1$, and evaluate all the formulae at leading order in 
the parameters $\delta_{i a}$. But we will instead work exactly in the 
quantities $\eta_{xa}$. In this situation, the would-be goldstone bosons that are absorbed 
by the gauge vectors are mostly linear combinations of the fields $M^x$, with only a small 
mixture of the fields $M^i$. We will also consider again the minimal situation where the 
number of gauge symmetry breaking fields $M^x$ with non-zero shifts $\eta_{x a}$ 
is equal to the number of $U(1)$ vector fields $V^a$. The shift matrix $\eta_{x a}$ is 
then a square matrix that admits an inverse $\eta_{a x}^{-1}$. 

In this case, we will take as starting point the following general form of the K\"ahler 
potential, with an unspecified non-canonical term for the Higgs fields:
\bea
\a\a K = \hat K \Big(M^i + \bar M^i - \sum_a \delta_{i a} V^a\Big) 
+ \tilde K \Big(M^x + \bar M^x - \sum_a \eta_{x a} V^a\Big) \nn \\
\a\a \hspace{26pt} + \sum_\alpha K_{\alpha \bar \alpha} \Big(M^i + \bar M^i - \sum_a \delta_{i a} V^a\Big) \, 
\bar Q^\alpha e^{\sum_a \! q_{\alpha a} V^a} \! Q^\alpha + \dots \,.
\label{Knoncanonical}
\eea
The superpotential is instead left to be a generic gauge-invariant holomorphic function of the 
shifting moduli $M^i$ and $M^x$:
\bea
\a\a W = \hat W \Big(M^i - \sum_{a,x} \delta_{ia} \eta^{-1}_{ax} \, M^x\Big) + \dots \,.
\eea
The dots denote as before possible additional terms which are of higher order in the fields and can 
therefore be neglected.

On the vacuum, the K\"ahler metric is in this case block-diagonal, and the soft masses take the 
same form as before:
\bea
\a\a m_{\alpha \bar \alpha}^2 = - g_{\alpha \bar \alpha} 
\bigg[ \partial_p \partial_{\bar q} \, {\rm ln}\, \Big(K_{\alpha \bar \alpha} e^{- \hat K/3}\Big) F^p F^{\bar q} 
+ \sum_a q_{\alpha a} D^a \bigg] \,.
\label{softmass2}
\eea
Our main task is again to compute the $D^a$'s in terms of the $F^i$'s, after exploiting the fact that 
in the minimal situation under consideration, gauge invariance completely fixes the VEV of 
$M^x$ and $F^x$ in terms of the VEV of $M^i$ and $F^i$.

\subsection{Relation between auxiliary fields}

The vector auxiliary fields is determined by the relation $D_a = - G_a$ and takes in this case the 
following form:
\bea
\a\a D_a = \sum_k \delta_{k a} \, \hat K_{k} + \sum_{x} \eta_{x a}\, \tilde K_x \,.
\label{formD2}
\eea
The mass matrix $M_{ab}^2$ of the vectors fields is given by
\bea
\a\a M_{ab}^2 = 2 \sum_{i,j} \delta_{i a} \delta_{\barj b} \, \hat K_{i \barj}
+ 2 \sum_{x,y} \eta_{x a} \eta_{y b} \, \tilde K_{xy} \,.
\label{formM2}
\eea

At the stationary point of the potential, the VEV that the $D_a$ actually develop 
are given by the relation (\ref{dynnew}):
\bea
\a\a D^a = - 2 \sum_b M^{-2ab} \Big(G_{b i \barj} \, F^i F^{\barj}  
+ G_{b x \bar y} \, F^x F^{\bar y} \Big) \,.
\label{expD2}
\eea
One easily calculates $G_{b i \barj} =  - \sum_k \delta_{k b} \hat K_{k i \barj}$
and $G_{b x \bar y} = - \sum_z \eta_{z b} \tilde K_{z x \bar y}$.
Substituting these expressions into eq.~(\ref{expD2}), one finds:
\bea
\a\a D^a = 2 \sum_b M^{-2ab} \bigg[
\sum_k \delta_{k b} \, \hat K_{i \barj k}\, F^i F^{\barj}
+ \sum_z \eta_{z b} \, \tilde K_{x \bar y z}\, F^x F^{\bar y}
\bigg] \,,
\label{evalDnew2}
\eea

Before proceeding, we shall now make the following further mild assumptions 
concerning the form of the K\"ahler potential of the Higgs fields:
\bea
\a\a \tilde K''=\;\mbox{diagonal} \,, 
\label{assum1}\\[1mm]
\a\a \tilde K''/(\tilde K')^2 \gg \hat K''/(\hat K')^2 
\label{assum2} \,.
\eea
These assumptions are not strictly necessary, but they simplify the computation 
enough to be able to carry it out in general.

The VEV of the $M^x$ are related to the VEV of the $M^i$. Indeed, the approximate
$D$-flatness conditions $D_a \simeq 0$ imply that 
$\sum_{x} \eta_{x a} \, \tilde K_x \simeq - \sum_k \delta_{k a} \, \hat K_{k}$.
Thanks to (\ref{assum1}), one can then formally extract the value of the Higgs fields
in terms of the inverse function $\tilde K_x^{\rm inv}$ of the function $\tilde K_x$. One finds:
\bea
\a\a M^x + \bar M^x \simeq \tilde K_x^{\rm inv} \big(v_x^2 \big)\,,
\label{gHH2}
\eea
where now
\bea
\a\a v_x = \Big(\!-\! \sum_{k,a} \delta_{k a} \eta_{a x}^{-1} \hat K_k \Big)^{1/2}\,.
\label{vx2}
\eea
As a consequence of (\ref{assum2}), the mass of the vector fields (\ref{formM2}) is again 
dominated by the second term. The mass matrix and its inverse can then be approximately 
written as:
\bea
\a\a M_{ab}^2 \simeq 2 \sum_{x} \eta_{x a} \eta_{x b} \tilde K_{x \bar x} \,,\quad\;
M^{-2ab} \simeq \frac 12 \sum_x \eta_{a x}^{-1} \eta_{b x}^{-1} \tilde K_{x \bar x}^{-1} \,.
\label{Mapprox2}
\eea

The values of the auxiliary fields $F_x$ are similarly related to the values of the auxiliary fields 
$F_i$. Indeed, the approximate relation (\ref{kinnew}) implies that 
$\sum_x \eta_{x a} F_x \simeq - \sum_k \delta_{k a} F_k$.
One deduces then that:
\bea
\a\a F_x =  - \sum_{k,a} \delta_{ka} \eta_{ax}^{-1} F_k \,.
\label{Fx2}
\eea
In this case, it is trivial to derive the values of the auxiliary fields with upper indices, since the 
fields $M^x$ and $M^i$ do not mix in the K\"ahler potential. One simply has $F^x = g^{x \bar x} F_{\bar x}$
and $F^i = g^{i \barj} F_{\barj}$. This implies then that 
\bea
\a\a F^x =  - \tilde K_{x \bar x}^{-1} \sum_{k,a} \delta_{ka} \eta_{ax}^{-1} \hat K_{k \barj} F^{\barj} \,.
\label{Fxup2}
\eea
Finally, the above expression can also be rewritten in a more illuminating form. Indeed, 
by the definition of the inverse function appearing in (\ref{gHH2}) one has 
$(\tilde K_x^{\rm inv})' = \tilde K_{x \bar x}^{-1}$, and thus:
\bea
F^x = \partial_i \, {\tilde K_x^{\rm inv}} \Big(\!-\! \sum_{k,c} \delta_{k c} \eta_{c x}^{-1} \hat K_k \Big)
F^i \,.
\eea

We can now evaluate eq.~(\ref{evalDnew2}). Using the above results (\ref{gHH2}), (\ref{Mapprox2}) 
and (\ref{Fxup2}) for $M^x + \bar M^x$, $M^{-2ab}$ and $F^x$, one obtains:
\bea
\a\a D^a = \sum_{x} \eta_{a x}^{-1}  \bigg[
\sum_{k,b} \delta_{kb} \eta_{bx}^{-1} \tilde K_{x \bar x}^{-1} \hat K_{i \barj k} 
+\!\! \sum_{k,l,b,c} \! \delta_{kb} \eta_{bx}^{-1} \delta_{l c} \eta_{c x}^{-1}
\tilde K_{x \bar x x} \, \tilde K_{x \bar x}^{-3} \, 
\hat K_{k \barj} \, \hat K_{i \bar l} \bigg]
F^i F^{\barj}  \,. \hspace{20pt}
\eea
Recalling the expression (\ref{vx2}) for $v_x$, and noticing that by the definition 
of the inverse function appearing in (\ref{gHH2}) one has 
$(\tilde K_x^{\rm inv})' = \tilde K_{x \bar x}^{-1}$ and 
$(\tilde K_x^{\rm inv})'' = - \tilde K_{x x \bar x} \tilde K_{x \bar x}^{-3}$, 
this result can finally be rewritten simply as 
\bea
\a\a D^a = - \sum_x \eta_{a x}^{-1} \, \partial_i \partial_{\barj} \, 
\tilde K_x^{\rm inv} \Big(\!-\! \sum_{k,c} \delta_{k c} \eta_{c x}^{-1} \hat K_k \Big)
F^i F^{\barj}  \,.
\label{resDa2}
\eea

The above result can again be derived in a much simpler way by integrating out the 
heavy vector fields at the superfield level. To do so, we define the following new 
gauge-invariant combinations of chiral and vector fields:
\bea
\a\a Q^{\prime \alpha} = Q^\alpha \exp \bigg\{\sum_{x,a} q_{\alpha a} \eta_{a x}^{-1} \, M^x \bigg\} \,, \nn \\
\a\a M^{\prime i} = M^i - \sum_{x,a} \delta_{i a} \eta_{a x}^{-1} \, M^x  \,, \\
\a\a V^{\prime a} = V^a - \sum_x \eta_{a x}^{-1} \Big(M^x + \bar M^x \Big) \nn \,.
\eea
These combinations of fields allow to rewrite the theory with one less type of fields, 
without any $M^{\prime x}$. This procedure can as before be interpreted as a gauge fixing 
corresponding to setting $M^{\prime x} = 0$, with the choice $\Lambda^a = -\,i \sum_x \eta_{a x}^{-1} \, M^x$ 
for the superfield gauge parameter. The K\"ahler potential can then be rewritten as
\bea
\a\a K = \hat K \Big(M^{\prime i} + \bar M^{\prime i} - \sum_a \delta_{i a} V^{\prime a} \Big)
+ \tilde K \Big(\!-\! \sum_a \eta_{x a} V^{\prime a}\Big) \nn \\
\a\a \hspace{26pt} 
+\, \sum_\alpha K_{\alpha \bar \alpha} \Big(M^{\prime i} + \bar M^{\prime i} - \sum_a \delta_{i a} V^{\prime a}\Big) \, 
\bar Q^{\prime \alpha} e^{\sum_a \! q_{\alpha a} V^{\prime a}} \! Q^{\prime \alpha} + \dots \,,
\eea
and the superpotential as
\bea
W = \hat W \Big(M^{\prime i}\Big) + \dots \,.
\eea
The superfield equations of motion for $V^{\prime a}$ are now easily computed by taking a derivative
with respect to $V^{\prime a}$. At leading order in the shift parameters $\delta_{i a}$, one finds just:
\bea
\a\a \sum_x \eta_{x a} \tilde K_x \Big(\!-\! \sum_b \eta_{x b} V^{\prime b}\Big) =  - \sum_k \delta_{k a} \hat K_k
\Big(M^{\prime i} + \bar M^{\prime i} \Big) \,.
\eea
From this one easily extracts the following solution for $\hat V^a$:
\bea
V^{\prime a} = - \sum_{x} \eta_{a x}^{-1} \, \tilde K_x^{\rm inv} \Big(\!-\! \sum_{k,c} \delta_{k c} \eta_{c x}^{-1} \, \hat K_k \Big) \,.
\label{solVa2}
\eea
Taking the $D$-component of (\ref{solVa2}), one finally recovers eq.~(\ref{resDa2}):
\bea
D^{\prime a} = - \sum_{x} \eta_{a x}^{-1} \, \partial_i \partial_{\barj} \, \tilde K_x^{\rm inv}
\Big(\!-\!\sum_{k,c} \delta_{k c} \eta_{c x}^{-1} \, \hat K_k \Big) F^{\prime i} F^{\prime \barj}\,.
\eea

\subsection{Structure of soft masses}

The soft masses of the visible fields are in this case given by the formulae (\ref{softmass2})
and (\ref{resDa2}), and have the following structure:
\bea
\a\a m_{\alpha \bar \alpha}^2 = - g_{\alpha \bar \alpha} \, \partial_i \partial_{\bar j}
\bigg[{\rm ln}\, \Big(K_{\alpha \bar \alpha} e^{- \hat K/3} \Big)
- \sum_{a,x} q_{\alpha a} \eta_{a x}^{-1}  \, 
\tilde K_x^{\rm inv} \Big(\!-\! \sum_{k,c} \delta_{k c} \eta_{c x}^{-1} \hat K_k \Big)
\bigg] F^i F^{\barj}\,. \hspace{10pt}
\label{softnew} 
\eea
The expression (\ref{softnew}) represents the generalization of (\ref{soft}) to the 
case where the Higgs fields have an arbitrary K\"ahler potential but do not directly mix
to the moduli fields. Taking $\tilde K(x) = e^x$, so that $\tilde K_x^{\rm inv}(y) = {\rm ln}(y)$,
one recovers the situation of previous section in the particular case $\tilde K_{xx} = 1$, 
after identifying $H^x = e^{M^x}$ and $q_{x a} = - \eta_{xa}$. Eq.~(\ref{softnew}) shows 
that in the situation where the Higgs fields have a non-canonical K\"ahler potential, the 
$D$-term contribution to soft masses has a functional dependence that is determined 
by the inverse function of the first derivative of this potential. This can in particular happen 
to vanish if the functional form of the K\"ahler potential for the Higgs fields is related to that 
of the moduli fields.

The physical soft masses, obtained by rescaling the above expression by the 
metric $g_{\alpha \bar \alpha}$, can be rewritten in the form:
\bea
\a\a m_\alpha^2 = m_{F \alpha}^{2} + \sum_{a,x} q_{\alpha a} \eta_{a x}^{-1} \, \Big(m_{F x}^2 - m_{Ax}^2\Big) \,,
\eea
where now
\bea
\a\a m_{F \alpha}^2 = - \partial_i \partial_{\barj}\,  {\rm ln} \Big(K_{\alpha \bar \alpha} e^{- \hat K/3}\Big) 
F^i F^{\barj}\,, \nn \\
\a\a m_{F x}^2 = - \partial_i \partial_{\barj}\, \tilde K_x^{\rm inv} \Big(e^{- \hat K/3} \Big)
F^i F^{\barj} \,, \\
\a\a m_{Ax}^2 = - \partial_i \partial_{\barj}\, \tilde K_x^{\rm inv} 
\Big(\!-\! \sum_c \eta^{-1}_{c x} \hat K_{c} e^{- \hat K/3} \Big) 
F^i F^{\barj} \,. \nn
\eea
The first term is the usual $F$-term contribution. The second term isolates as before a general part of the 
$D$ term contribution consisting of a linear combination of generalized $F$-term masses of the Higgs fields 
suitably weighted by their charges. Finally, the third contributions encodes the non-trivial effects related to 
anomalies.

The flatness and stability constraints (\ref{flatnessnew}) and (\ref{stabilitynew}) are as 
before dominated by the moduli fields. The effects of the Higgs fields and of the vector 
fields are again small as a consequence of the assumption that the shift vectors $\delta_{ia}$ 
of the supersymmetry breaking fields are small. In the end, the flatness and stability constraints 
read thus simply:
\bea
\a\a g_{i \barj} \, F^i F^{\barj} \simeq 3\, m_{3/2}^2 \,, \\[1mm]
\a\a R_{i \barj p \bar q} \, F^i F^{\barj} F^p F^{\bar q} \lsim 6\, m_{3/2}^4 \,.
\eea

As before, the physical components of the Goldstino direction are obtained by rescaling the
auxiliary fields by the square root of the metric, and the flatness and stability constraints are
unchanged.

In the frame of normal coordinates around the vacuum, corresponding to canonically normalized 
fields, the physical soft scalar masses can also be written as
\bea
\a\a m_{\alphacap \bar \alphacap}^2 = \bigg[\frac 13 \, \delta_{I \bar J}
- R_{\alphacap \bar \alphacap I \bar J} +\, \sum_{A,X} q_{\alphacap A} \eta_{A X}^{-1}  \, S_{X \bar X I \bar J}
\bigg] F^I F^{\bar J}\,, \hspace{10pt}
\label{softcan2} 
\eea
and the flatness and stability conditions read
\bea
\a\a \delta_{I \bar J} \, F^I \bar F^{\bar J} = 3\, m_{3/2}^2 \,,\;\; \\
\a\a R_{I \bar J P \bar Q} \, F^I \bar F^{\bar J} F^P \bar F^{\bar Q} \le 6\, m_{3/2}^4 \,,
\eea
where $q_{\alphacap A} \equiv q_{\alpha a}$, $\eta_{X A} \equiv \eta_{x a}$ and
\bea
\a\a R_{I \bar J P \bar Q} = e_I^i e_{\bar J}^{\barj} e_P^p e_{\bar Q}^{\bar q} 
\Big(\hat K_{i \barj p \bar q} - \hat K^{r \bar s} \hat K_{i p \bar s} \hat K_{\barj \bar q r} \Big) \,, \nn \\
\a\a R_{\alphacap \bar \alphacap I \bar J} = e_I^i e_{\bar J}^{\barj} \, 
\partial_i \partial_{\barj} \, {\rm ln} \, \Big(K_{\alpha \bar \alpha} \Big) \,, \label{curvflat2} \\
\a\a S_{X \bar X I \bar J} = e_I^i e_{\bar J}^{\barj} \, \partial_i \partial_{\barj} \, \tilde K_x^{\rm inv}
\Big(\!-\!\sum_{k,c} \delta_{k c} \eta_{c x}^{-1} \hat K_k \Big) \,. \nn
\eea

\subsection{Basic moduli}

Let us finally see what happens in the particular case where the supersymmetry breaking fields
and the gauge symmetry breaking fields both belong to a set of universal string moduli, each 
spanning an $SU(1,1)/U(1)$ manifold. In general, all the supersymmetry breaking and gauge symmetry 
breaking fields can shift under the $U(1)$ symmetries, with shifts $\delta_{ia}$ and $\eta_{xa}$. 
The K\"ahler potential has the following structure:
\bea
\a\a \hat K = - \sum_i \,c_i\,{\rm ln}\,\Big(M^i + \bar M^i - \sum_a \delta_{ia} V^a \Big) \,,
\label{KpotMi} \\
\a\a \tilde K = - \sum_x \,c_x\,{\rm ln}\,\Big(M^x + \bar M^x - \sum_a \eta_{xa} V^a \Big) \,.
\label{KpotMx}
\eea
In this interesting situation, the approximations done to derive eq.~(\ref{softcan2}) are unfortunately 
not valid. In particular, one has $\tilde K''/(\tilde K')^2 \sim \hat K''/(\hat K')^2$, in conflict with 
(\ref{assum2}). As a consequence, it is not possible to neglect terms with higher powers 
of $\delta_{ia}$, and one has to do an algebraically exact computation. 
This is feasible, thanks to the simple form of the K\"ahler potential, but 
unfortunately only a case by case analysis seems to be possible. Notice nevertheless 
that applying eqs.~(\ref{curvflat2}), one would find:
\bea
\a\a R_{I \bar J P \bar Q} = \frac 2{c_I} \, \delta_{I \bar J P \bar Q} \,, \label {Rconst2} \\[1mm]
\a\a S_{X \bar X I \bar J} = - \frac {2\, c_X}{c_I} \, \frac {t_{IX}}{\big(\sum_K t_{KX}\big)^2} \, \delta_{I \bar J} 
+ \frac {2\, c_X}{\sqrt{c_I c_J}} \, \frac {t_{IX}\,t_{J X}}{\big(\sum_K t_{KX}\big)^3}
\label{Snonconst2} \,,
\eea
in terms of the quantities
\bea
t_{I X} = - \sum_a \delta_{ia} \eta_{ax}^{-1} c_i \big(M^i + \bar M^i \big)^{-1} \,.
\label{Zdef2}
\eea
The quantity $S_{X \bar X I \bar J}$ has again a fixed dependence on the moduli 
fields. This is non-trivial when several supersymmetry breaking fields shift under 
$U(1)$ symmetries. But whenever there is a single shifting modulus $M^0$ breaking 
supersymmetry and being stabilized independently, one has $t_{IX} = t_{0X}\,\delta_{0I}$ 
and the result vanishes: $S_{X \bar X I \bar J} = 0$. This suggests that in this case 
there is no $D$-term effect in the soft masses.

For any fixed number of supersymmetry and gauge symmetry breaking fields with 
K\"ahler potential given by (\ref{KpotMi}) and (\ref{KpotMx}), it is possible to compute the exact
expression for the $D$-term contribution to soft masses in a rather straightforward way, by 
using the method of integrating out the vector multiplets at the superfield level. In the presence of 
a single supersymmetry breaking field $M^0$ with non-trivial shifts $\delta_{0a}$, the equations 
of motions of the vector superfields imply that $\sum_a Y_{x a} V^{\prime a} = M^{\prime 0} + \bar M^{\prime 0}$,
in terms of the square matrix $Y_{xa} = \delta_{0a} + (c_0/c_x) \sum_b \delta_{0b} \, \eta^{-1}_{bx} \eta_{xa}$. 
The solution of this linear system of equations takes then the simple form $V^{\prime a} = \big(\sum_x Y_{ax}^{-1}\big)
\big(M^{\prime 0} + \bar M^{\prime 0} \big)$. This result implies that the corresponding $D^a$ 
vanish, and shows that there is indeed no effect in this particular case. It should also be noted 
that in this case the VEV of the Higgs scalar fields and their auxiliary partners cannot be neglected
and contribute on the same footing as the moduli fields to the vacuum energy and the moduli 
soft masses, and thus to the flatness and stability conditions. In other words, one must 
in this case include in the set of multiplets contributing to supersymmetry breaking also the Higgs 
multiplets, and since the VEV of these multiplets are determined in terms of the VEV of the other
moduli multiplets, a special restriction on the Goldstino direction arises.

\section{Toroidal string models}
\setcounter{equation}{0}

Let us now study more in detail what happens in the simplest string models, based on a 
toroidal compactification geometry with some discrete orbifold or orientifold projection,
under the assumption that the supersymmetry breaking sector is identified with the 
sector of the untwisted moduli. These include the dilaton $S$, the K\"ahler class and complex 
structure moduli $T_r$ and $U_s$, and the Wilson line moduli $Z_a$. The scalar fields of the 
corresponding chiral multiplets control respectively the string coupling, the size and the shape 
of the compactification geometry, and the structure of the gauge bundle. The K\"ahler potential 
of such moduli is determined, at leading order in the weak-coupling and low-energy expansions, 
by a simple dimensional reduction of the minimal ten-dimensional supergravity theory. On a torus, 
the moduli space would be the product of an $SU(1,1)/U(1)$ factor, parametrized 
by $S$, and an $SO(6,6+r)/(SO(6) \times SO(6+r))$ factor, $r$ being the rank of the gauge group, 
parametrized by the $T_r$, $U_s$ and $Z_a$. The discrete projection defining the model 
preserves the first factor but reduces the second factor to a submanifold. This is in general
a coset K\"ahler manifold which cannot be completely factorized. More precisely, it consists 
of up to $6$ basic dimensions spanning $SU(1,1)/U(1,1)$ submanifolds, associated with the 
universal K\"ahler class and complex structure moduli $T_r$ and $U_s$, and a variable number 
of additional dimensions, associated with additional model-dependent moduli, which enhanced the 
product of these basic factors to a larger and more symmetric manifold. For simplicity, we will assume 
that only the basic "diagonal" moduli associated with the $SU(1,1)/U(1)$ submanifolds participate 
to supersymmetry breaking, whereas the additional "non-diagonal" moduli enhancing the scalar
manifold do not have a relevant breaking effect. This assumption is not expected to represent a 
severe limitation, because the latter are associated with additional isometries, which suggest that
it should be possible to rephrase in an equivalent way effects due to off-diagonal moduli as a 
effects due to diagonal moduli \footnote{This is certainly true for the masses of the moduli themselves, 
and the off-diagonal fields do not allow to alleviate the flatness and stability conditions arising for 
the basic moduli \cite{grs2}. It is generically expected to hold true also for the soft masses, provided 
however that the isometry group of the moduli space survives as a global symmetry for the whole 
scalar manifold, including the matter fields.}.

Ignoring for the moment vector fields, the K\"ahler potential for the moduli sector has the following 
simple and separable structure, at leading order in the low-energy and week-coupling expansions
\cite{STU1,STU2,STU3}:
\bea
\a\a \hat K = - {\rm ln} \Big(S + \bar S\Big) 
- \sum_r  {\rm ln} \Big(T_r + \bar T_r \Big)
- \sum_s  {\rm ln} \Big(U_s + \bar U_s \Big) \,.
\eea
The flatness condition implies that the rescaled auxiliary fields can be parametrized in terms 
of a Goldstino angle $\theta$, controlling the relative importance of the dilaton $S$ and the geometric 
moduli $T_r$ and $U_s$, some spherical parameters $\Theta_r$ and $\Theta_s$, satisfying 
the constraint $\sum_r \Theta_r^2 + \sum_r \Theta_s^2 = 1$ and parametrizing the relative 
importance of the different geometric moduli, as well as some arbitrary phases $\gamma$, $\gamma_r$ 
and $\gamma_s$ (the bars denote flat indices) \cite{BIM}:
\bea
\a\a F^{\bar S} = \sqrt{3}\,e^{i \gamma}\sin \theta \, m_{3/2} \,, 
\label{FS}\\[1mm]
\a\a F^{\bar T_r} = \sqrt{3}\, \Theta_r \, e^{i \gamma_r} \cos \theta \, m_{3/2} \,,\quad\;
F^{\bar U_s} = \sqrt{3}\, \Theta_s \,e^{i \gamma_s} \cos \theta \, m_{3/2}  
\label{FTFU}\,.
\eea
The stability condition implies further restrictions on the parameters $\theta$ and $\Theta_r$. 
More precisely, it reads 
\bea
\a\a \sin^4 \theta + \Big(\sum_r \Theta_r^4 + \sum_s \Theta_s^4\Big) \cos^4 \theta \lsim \frac 13 \,.
\label{metamod}
\eea 
Notice that the quantity $z = \sum_r \Theta_r^4 + \sum_s \Theta_s^4$ is maximal when one of the 
$\Theta_{r,s}$ is equal to $1$ and all the other $0$, and minimal when all the $\Theta_{r,s}$ are
equal to $1/\sqrt{n}$, where $n$ is the total number of geometric moduli. So $z \in [1/n,1]$,
and small and large values of $z$ correspond respectively to very democratic and antidemocratic
distributions of the breaking among the geometric moduli. It is straightforward to see that the condition
(\ref{metamod}) can be satisfied for some $\theta$ only if $z \lsim 1/2$. This means that the breaking must 
be distributed over the various geometric moduli in a sufficiently democratic way, in order to increase
their weight compared to the dilaton in the condition. Thus, one finds the necessary condition
\bea
\a\a \sum_r \Theta_r^4 + \sum_s \Theta_s^4 \lsim \frac 12 \,.
\eea
When this condition is satisfied, that is $z \lsim 1/2$, the range of Goldstino angles satisfying 
the bound is given by $\theta \in \big[\theta_-,\theta_+ \big]$, where 
$\theta_\pm = {\rm arccos} \big[\big(1 \pm \sqrt{(1- 2 z )/3}\big)/\big(1 + z\big)\big]^{1/2}$. 
For $z \simeq 1/2$, there is only one critical value $\theta_0 \simeq {\rm arcsin} \, (3^{-1/2})$. 
For $z < 1/2$, there are distinct minimal and maximal values $\theta_-$ and $\theta_+$ departing 
monotonically from the critical value $\theta_0$. For $z \simeq 1/3$, these reach the values 
$\theta_- \simeq 0$ and $\theta_{+} \simeq \pi/4$. For smaller $z \ll 1/3$, the minimal value stays 
unchanged, $\theta_{-} \simeq 0$, whereas the maximal value saturates at the the absolute upper 
bound $\theta_+ \simeq {\rm arcsin} \, (3^{-1/4})$. In any case, the Goldstino angle must therefore 
certainly satisfy the following bound:
\bea
\a\a 0 \le \sin^2 \theta <  \frac 1{\sqrt{3}} \,.
\label{constraintgold}
\eea
The above results show that none of the moduli is allowed to dominate supersymmetry breaking 
on its own \cite{grs1} (see also \cite{BdA}). In particular, dilaton domination is excluded. On the other 
hand, it is conceivable that the $3$ or more of the geometric moduli may dominate. 

In the presence of heavy vector fields associated to broken $U(1)$ symmetries, some or 
all of the moduli may acquire a small shift transformation law. A small VEV for $D$ is then 
generated. Its effect on the vacuum energy and the moduli masses can be neglected, and 
the above results concerning the Goldstino direction remain approximately valid. On the 
other hand, this small $D$ gives significant effects on the soft terms of all the charged fields, 
whose form depends on the details of the models.

\subsection{Heterotic models}

Let us first study the case of heterotic orbifold models \cite{HetOrb1,HetOrb2}. 
In these models, there can be only one anomalous $U(1)$ symmetry, 
and the Green--Schwarz mechanism involves only $S$ and not $T_r$ or $U_s$. 
The Higgs fields are always charged matter fields $H^x$ and have a canonical 
K\"ahler potential. 

The situation is of the type discussed in section 4, and the general form of the 
K\"ahler potential is given by eq.~(\ref{Kcanonical}). The potential of the moduli 
has the simple form:
\bea
\a\a \hat K = - {\rm ln} \Big(S + \bar S - \sum_a \delta_a V^a \, \Big) 
- \sum_r  {\rm ln} \Big(T_r + \bar T_r \Big)
- \sum_s  {\rm ln} \Big(U_s + \bar U_s \Big) \,.
\eea
The couplings between the moduli fields and the matter and Higgs fields also 
have in this case a simple structure and depend on constant modular weights 
$n_{\alpha r}$, $n_{\alpha s}$ and $n_{x r}$, $n_{x s}$, representing the charges 
of the matter field $Q_\alpha$ and the Higgs field $H^x$ with respect to the $U(1)$ 
isometry transformations associated to the moduli $T_r$, $U_s$ \cite{MetricHet,IbanezLust}:
\bea
\a\a K_{\alpha \bar \alpha} = \prod_r \Big(T_r + \bar T_r \Big)^{n_{\alpha r}}
\prod_s \Big(U_s + \bar U_s \Big)^{n_{\alpha s}} \,, \\
\a\a K_{x \bar x} = \prod_r \Big(T_r + \bar T_r \Big)^{n_{x r}}
\prod_s \Big(U_s + \bar U_s \Big)^{n_{\alpha s}} \,.
\eea
Notice that the corresponding mixed components of the Riemann tensor are given 
simply by $R_{\alphacap \bar \alphacap I \bar J} = - n_{\alpha i}\, \delta_{i \barj}$ 
and $R_{X \bar X I \bar J} = - n_{x i}\,\delta_{i \barj}$, and are thus field-independent 
and diagonal. This leads to a very simple structure for the $F$-term contribution 
to soft terms \cite{BIM}.

The expression for the physically normalized soft scalar masses can be obtained 
by applying the general results derived in section 4. One finds the following very 
simple result
\bea
\a\a \frac {m^{2}_{\alpha}}{m_{3/2}^2} =  
\Big[1 + n_{\alpha} \cos^2 \theta\Big] - \sum_{a,x} q_{\alpha a} q_{a x }^{-1} 
\Big[n_{x} \cos^2 \theta + 3 \, \sin^2 \theta \Big] \,,
\label{mhet}
\hspace{25pt}
\eea
in terms of the total modular weights
\bea
\a\a n_\alpha = 3\, \Big(\sum_r n_{\alpha r}  \Theta_r^2 + \sum_s n_{\alpha s} \Theta_s^2 \Big) \,, \\
\a\a n_x = 3\, \Big(\sum_r n_{x r}  \Theta_r^2 + \sum_s n_{x s} \Theta_s^2 \Big) \,.
\eea
In the particular case where a single anomalous $U(1)$ is present, without any extra 
non-anomalous $U(1)$, eq.~(\ref{mhet}) simplifies to
\footnote{The expression (\ref{mhet1}) differs from the one derived in 
\cite{KawamuraD1, KawamuraD2,KawamuraD3} by the coefficient and the dependence on the 
Goldstino angle of the last term, which plays an important role for the issue of flavor universality. 
This can be traced back to the fact that the expression obtained by applying eq.~(\ref{FxHx}) 
for the values of the auxiliary components of the Higgs fields differs from the one used in 
\cite{KawamuraD1,KawamuraD2,KawamuraD2}, where the dependence on $H^x$ implied 
by gauge invariance in the superpotential stabilizing $S$ was not taken into account.}
\bea
\a\a \frac {m^2_{\alpha}}{m_{3/2}^2} =  
1 + \Big(n_\alpha - \frac {q_\alpha}{q_x} \, n_x \Big) \cos^2 \theta
- 3 \, \frac {q_\alpha}{q_x} \sin^2 \theta \,.
\label{mhet1}
\hspace{25pt}
\eea

The formula (\ref{mhet1}) shows that the soft scalar masses are universal 
for an arbitrary $\theta$ only if both the modular weights $n_\alpha$ 
and the $U(1)$ charges $q_\alpha$ are universal. In the dilaton domination
limit $\theta \to \pi/2$, it is enough that the charges $q_\alpha$ are universal.
This situation is however excluded by metastability considerations.
On the other hand, in the geometric moduli domination limit $\theta \to 0$, 
it is enough that the combination $n_\alpha - q_\alpha/q_x \, n_x$, acting as an effective 
modular weight, is universal. This situation is allowed by metastability 
consideration, and we will consider it as our preferred scenario.
Summarizing, one could conceivably achieve universal soft masses if 
$\theta \to 0$ and the differences between the modular weights and the charges 
of any pair of different flavors of matter fields satisfy:
\bea
\Delta n - \frac {\Delta q}{q_x} \, n_x = 0 \,.
\label{relation}
\eea

It is now worth to comment on what happens in the situation of \cite{Dudas1,Dudas2},
where the $U(1)$ symmetries play the role of flavor symmetries. 
In these works it was shown that a non-zero perturbative Yukawa coupling mixing 
two flavors can arise from a gauge-invariant and modular-invariant higher-dimensional 
operator involving powers of the gauge symmetry breaking fields $H^x$, provided that the 
differences between the modular weights and the charges of the two involved matter fields 
satisfy a selection rule which is precisely given by eq.~(\ref{relation}). In such a situation, the 
contribution to the soft scalar masses that we have computed above would display an 
improved structure, with the difference between two different flavors given simply 
by $\Delta m^2/m_{3/2}^2 = - 3 \, \Delta n/n_x \sin^2 \theta$ 
\footnote{This result differs qualitatively from the one derived \cite{Dudas1,Dudas2},
as far as the dependence on the Goldstino angle is concerned. We believe that this
is due to the fact that the simple minimization procedure used in \cite{Dudas1,Dudas2}
to derive the value of the vector auxiliary field also implicitly neglects the fact that gauge 
invariance implies a non-trivial dependence of the $F$-term potential on the Higgs fields,
which does not amount just to their soft mass terms.}.
In the volume domination limit $\theta \to 0$, one finds in particular universal diagonal 
values for the soft scalar mass matrix. However, it must be emphasized that the relation 
(\ref{relation}) also allows the appearance of gauge-invariant and modular-invariant 
higher-dimensional operators in the K\"ahler potential mixing two different families of 
matter fields and powers of the gauge symmetry breaking fields $H^x$. These would 
induce subleading off-diagonal elements for the soft mass matrix, which are as dangerous 
as differences between diagonal elements, and represent the main difficulty in constructing
viable supersymmetric flavor models. It would nevertheless be interesting to further explore 
whether this somewhat peculiar situation can help in getting a simple and satisfactory 
supersymmetric flavor model in this framework, with the anomalous $U(1)$ symmetry,
possibly together with other non-anomalous but Higgsed $U(1)$ symmetries, playing the 
role of flavor symmetries.

\subsection{Brane models}

Let us consider now the case of intersecting brane models \cite{InterBrane,SusyInt1,SusyInt2}. 
In these models, there can be several anomalous $U(1)$'s, and the Green--Schwarz mechanism 
compensating them involves linear combinations of $S$ and the $U_s$, but not the $T_r$. In this 
case, the Higgs fields compensating the Fayet--Iliopoulos terms induced by the anomalous $U(1)$'s 
can be either additional matter fields $H^x$ with a canonical K\"ahler potential and a moduli-dependent 
wave-function factor, or the complex structure moduli $U_s$, which have a non-canonical K\"ahler potential
\footnote{It is useful to recall that the angles defining the geometric orientation of the branes are 
determined by the VEV and the shift charges of $S$ and $U_s$, whereas the nature of the 
intersection is possibly influenced by localized Higgs fields $H^x$ \cite{SusyInt1,SusyInt2,Cremades,VZ}. 
The condition for approximate supersymmetry that the sources induced for the $D$-terms by the 
fields $S$ and $U_s$ vanish translates then into the conditions that the relative angles between the 
branes should correspond to rotations belonging to an $SU(3)$ subgroup of $SO(6)$ and preserving 
a $U(1)$ factor. The two situations described above can then be interpreted as follows. In the first case, 
the values of the angles are fixed by some $F$-term dynamics and violate supersymmetry; the 
VEV of the additional Higgs fields can then be geometrically interpreted as corresponding to a 
recombination of the branes through their intersections to a final state with new angles preserving 
approximately supersymmetry. In the second case, the values of the angles are instead fixed by the 
$D$-term dynamics itself, and do thus approximately preserve supersymmetry from the beginning.}.

In the case where all the moduli $S$, $T_r$ and $U_s$ are stabilized by $F$-term effects and the 
$D$-terms are compensated by canonical matter Higgs fields $H^x$, the situation is of the type 
discussed in section 4, and the general form of the K\"ahler potential is given by eq.~(\ref{Kcanonical}). 
The potential of the moduli has the simple form:
\bea
\a\a \hat K = - {\rm ln} \Big(S + \bar S \! - \sum_a \delta_a V^a \Big) 
- \sum_r  {\rm ln} \Big(T_r + \bar T_r \Big)
- \sum_s  {\rm ln} \Big(U_s + \bar U_s \! - \! \sum_a \delta_{sa} V^a\Big) \,.
\eea
The form of the relevant couplings between the moduli fields and the matter and Higgs 
fields have been studied only very recently in the literature \cite{MetricInt1,MetricInt2}. 
Their precise form is still under debate \cite{ReviewInt}, and does not seem to rest on 
any symmetry argument. For this reason, we shall leave it arbitrary and take:
\bea
\a\a K_{\alpha \bar \alpha} =  K_{\alpha \bar \alpha}
\Big(T_r + \bar T_r, U_s + \bar U_s  \! - \! \sum_a \delta_{s a}V^a \Big)\,,\\
\a\a \tilde K_{x \bar x} = \tilde K_{x \bar x}
\Big(T_r + \bar T_r, U_s + \bar U_s  \! - \! \sum_a \delta_{s a}V^a \Big) \,.
\eea
We can however still parametrize the corresponding mixed components of the Riemann 
tensor as $R_{\alphacap \bar \alphacap I \bar J} = - n_{\alpha i \barj}$ and 
$R_{X \bar X I \bar J} = - n_{x i \barj}$. The quantities $n_{\alpha i \barj}$ 
and $n_{x i \barj}$ play then the role of effective modular weights. In general they 
can depend on the moduli fields and have non-vanishing off-diagonal 
components. The structure of the $F$-term contribution to soft terms is 
consequently more complicated in this case \cite{SoftInt1,SoftInt2}. 

We can now apply the results obtained in section 4 to derive the physically
normalized soft masses. The result can be written in terms of the above-defined 
effective modular weights and reads
\bea
\a\a \frac {m^2_{\alpha}}{m^2_{3/2}} = 
\Big[1 + n_{\alpha} \cos^2 \theta\Big] - \sum_{a,x} q_{\alpha a} q_{a x }^{-1} 
\Big[\big(n_x + b_x \big) \cos^2 \theta + a_x \sin^2 \theta + c_x \sin 2\theta \Big] \,,
\label{mbrane}
\hspace{25pt}
\eea
in terms of the total modular weights
\bea
\a\a n_\alpha = 3\, \Big(\sum_{r,r'} n_{\alpha r r'}  \Theta_r \Theta_{r'} 
+ \sum_{s,s'} n_{\alpha s s'} \Theta_s \Theta_{s'} \Big) \,, \\
\a\a n_x = 3\, \Big(\sum_{r,r'} n_{x r r'}  \Theta_r \Theta_{r'} 
+ \sum_{s,s'} n_{x s s'} \Theta_s \Theta_{s'} \Big) \,.
\eea
and the following functions of the variables (\ref{tvar}) depending on the inverse VEV
of the moduli fields:
\bea
\a\a a_x = 6 \, \frac {t_{S x}}{t_{S x} + \sum_s t_{U_s x}} 
- 3\, \Big(\frac {t_{S x}}{t_{S x} + \sum_s t_{U_s x}} \Big)^2 \,, \nn \\[1mm]
\a\a b_x = 6 \sum_s \frac {t_{U_s x}}{t_{S x} + \sum_s t_{U_s x}} \, \Theta_s^2 
- 3\, \Big(\sum_{s}  \frac {t_{U_s x}}{t_{S x} + \sum_s t_{U_s x}} \, \Theta_s \Big)^2 \,, \\
\a\a c_x = -\, 3\, \Big(\frac {t_{S x}}{t_{S x} + \sum_s t_{U_s x}} \Big) 
\Big(\sum_{s}  \frac {t_{U_s x}}{t_{S x} + \sum_s t_{U_s x}} \, \Theta_s  
\cos \big(\gamma_s - \gamma \big) \Big)\,. \nn
\eea
In the particular case where a single anomalous $U(1)$ is present, without 
any additional anomalous or non-anomalous $U(1)$ and only one Higgs field,  
the above expression simplifies and yields
\bea
\a\a \frac {m^2_{\alpha}}{m^2_{3/2}}  = 
1 + \Big(n_\alpha - \frac {q_\alpha}{q_x} \, \big(n_x + b_x\big) \Big) \cos^2 \theta
- \frac {q_\alpha}{q_x} \, \Big(a_x \sin^2 \theta + c_x \sin 2\theta \Big)\,.
\label{mbrane1}
\hspace{25pt}
\eea

The formula (\ref{mbrane1}) shows that the soft scalar masses are as before 
universal for an arbitrary $\theta$ only if both the modular weights $n_\alpha$ 
and the $U(1)$ charges $q_{\alpha}$ are universal. In the dilaton domination
limit $\theta \to \pi/2$, it is enough that the charges $q_\alpha$ are universal,
but this situation is again excluded by metastability considerations. In the 
geometric moduli domination limit $\theta \to 0$, on the other hand, it is enough 
that the corrected modular weight $n_\alpha - q_{\alpha}/ q_{x} \, \big(n_x + b_x \big)$ 
is universal. Since $n_{\alpha}$ and $n_{x}$ may have related functional forms, 
but $b_x$ is instead an unrelated function, this universality could plausibly arise 
only in situations where $b_x = 0$. This happens for instance in the case where 
all the complex structure moduli are stabilized in a supersymmetry way, so that 
$\Theta_s = 0$. Summarizing, one could conceivably achieve universal soft masses 
if $\theta \to 0$, $\Theta_s \to 0$, and the differences between the effective 
modular weights and the charges of any pair of different flavors of matter fields 
satisfy:
\bea
\Delta n - \frac {\Delta q}{q_x} \, n_x = 0 \,.
\label{relationbis}
\eea
In this situation, supersymmetry would be dominated only by the K\"ahler moduli.
Since there are $3$ of them, the metastability condition is then marginally violated, 
and subleading corrections to the K\"ahler potential are expected to play a crucial 
role in the stabilization of the lightest modulus \cite{grs1}.

In the case where only the $S$ and $T_r$ moduli are stabilized by $F$-term effects, 
whereas the $U_s$ moduli are stabilized by $D$-terms and act as Higgs fields, the 
situation is that of section 5, and the general form of the K\"ahler potential is given 
by eq.~(\ref{Knoncanonical}). The potential of the supersymmetry breaking moduli 
and the Higgs fields have the forms:
\bea
\a\a \hat K = - {\rm ln} \Big(S + \bar S - \sum_a \delta_a V^a \Big) 
- \sum_r  {\rm ln} \Big(T_r + \bar T_r \Big) \,, \\
\a\a \tilde K = - \sum_s  {\rm ln} \Big(U_s + \bar U_s - \sum_a \eta_{sa} V^a\Big) \,.
\eea
The coupling between the moduli fields and the matter fields is again left arbitrary,
whereas the coupling between the moduli fields and the Higgs fields is in this case 
absent:
\bea
\a\a K_{\alpha \bar \alpha} =  K_{\alpha \bar \alpha}
\Big(T_r + \bar T_r, U_s + \bar U_s  - \sum_a \eta_{s a}V^a \Big) \,.
\eea
As before, we can still parametrize the mixed components of the Riemann tensor 
as $R_{\alphacap \bar \alphacap I \bar J} = - n_{\alpha i \barj}$, where 
$n_{\alpha i \barj}$ is some effective modular weight depending 
on the moduli fields and possessing non-vanishing off-diagonal entries. 
On the other hand, $R_{X \bar X I \bar J} = 0$ in this case.

We can now apply the results obtained in section 5 to derive the physically
normalized soft masses. As discussed in section 5, the $D$-term 
contribution to the soft masses vanishes in this case. One finds then simply:
\bea
\a\a \frac {m^2_{\alpha}}{m_{3/2}^2} =  
\Big[1 + n_{\alpha} \cos^2 \theta\Big] \,,
\label{mbrane2}
\hspace{25pt}
\eea
in terms of the total modular weights
\bea
\a\a n_\alpha = 3\, \Big(\sum_{r,r'} n_{\alpha r r'}  \Theta_r \Theta_{r'} 
+ \sum_{s,s'} n_{\alpha s s'} \Theta_s \Theta_{s'} \Big) \,.
\eea
This result is in particular valid in the minimal situation involving three anomalous
$U(1)$, whose $D$-terms stabilize the three complex structure moduli $U_s$ relative
to the dilaton $S$.

As mentioned at the end of section 5, there is in this case a restriction on the Goldstino 
direction. Indeed, the bosonic components of the $3$ complex structure multiplets 
$U_s$ are completely determined in terms of the those of the dilaton $S$. More 
precisely, from the $3$ $D$-flatness conditions one finds the relations 
$\big(U_s + \bar U_s\big)^{-1} \! = \big(\sum_a \! Z_{sa}^{-1}\big) \big(S + \bar S\big)^{-1}\!$ 
and $F_{U_s} = \big(\sum_a \! Z_{sa}^{-1}\big) F_S$, in terms of the $3$ by $3$ matrix 
$Z_{sa} = - \eta_{s a}/\delta_a$. This implies then that the rescaled auxiliary fields 
satisfy the simple relation $F^{\bar U_s} = F^{\bar S}$. Comparing with the general
parametrization of eqs. (\ref{FS}) and (\ref{FTFU}), this implies the restriction 
\bea
\a\a \Theta_s = \tan \theta \,.
\label{restriction}
\eea

The formula (\ref{mbrane2}) shows that the soft scalar masses are in this case 
insensitive to the $U(1)$ charges $q_{\alpha a}$, and are universal for an arbitrary 
$\theta$ if the modular weights $n_\alpha$ are universal. Notice finally that 
in the geometric moduli domination limit $\theta \to 0$ nothing really special 
happens for the soft masses, but the relation (\ref{restriction}) implies then that 
the complex structure moduli give a negligible effect as well and only the K\"ahler 
moduli break supersymmetry. Since there are $3$ of these, the metastability 
condition is then again marginally violated, implying that subleading corrections 
to the K\"ahler potential are expected to play a crucial role in stabilizing at least 
one of them \cite{grs1}.

\section{Conclusions}
\setcounter{equation}{0}

In this paper, we have studied in some generality the structure of soft scalar masses 
in superstring models with heavy anomalous $U(1)$ vector fields. We have considered 
the minimal situation where the sources induced by the supersymmetry breaking 
fields in the $D$-terms are approximately compensated, at leading order in $m_{3/2}$, 
by some Higgs fields taking suitable VEV. We have computed with two different 
methods the structure of the $D$-term contribution to soft scalar masses, relative 
to the usual $F$-term effect. We have shown that the result significantly depends 
on the functional form of the K\"ahler potential for the Higgs fields. In particular, 
assuming that the supersymmetry breaking fields are untwisted moduli fields with 
a logarithmic K\"ahler potential, we find that the effect of heavy vector fields is non-trivial
when the Higgs fields are matter fields with a quadratic K\"ahler potential, but 
can accidentally vanish if they happen to be moduli fields with a logarithmic 
K\"ahler potential.

For heterotic orbifold models, we find a result for the soft scalar masses that slightly 
differs from previous analyses. We assume that the supersymmetry breaking moduli 
$S$, $T_r$ and $U_s$ are stabilized by some $F$-term effects and that the $D$-terms 
are approximately compensated by a minimal set of matter Higgs fields $H^x$ with a 
quadratic K\"ahler potential. The resulting soft scalar masses have then a pretty simple 
structure, with a non-trivial $D$-term contribution that is as important as the $F$-term 
contribution. The total masses depend on the Goldstino direction of supersymmetry breaking, 
but not on the scalar VEV of the supersymmetry breaking fields. They can be flavor-universal 
independently of any further assumption about the superpotential and the Goldstino 
direction only under the very strong condition that both the modular weights $n_\alpha$ 
and the $U(1)$ charges $q_{\alpha a}$ of the matter fields are universal. On the other 
hand, in the geometric moduli domination limit, which on the contrary of the dilaton 
mediation limit is compatible with flatness and metastability of the vacuum, flavor 
universality of the soft masses is guaranteed under the milder requirement that suitable 
linear combinations of modular weights and $U(1)$ charges are universal. 

For intersecting brane models, on the other hand, we find new results displaying a 
richer variety of possibilities. In this case, there exist two qualitatively different options 
for the stabilization of the various fields, leading to radically different structures for the 
soft scalar masses. A first possibility is that all the supersymmetry breaking moduli
$S$, $T_r$ and $U_s$ are stabilized by $F$-term effects and that the $D$-terms are 
approximately compensated by a minimal set of matter Higgs fields $H^x$ with a quadratic 
K\"ahler potential, much as in heterotic models. The resulting soft scalar masses have then 
again a non-trivial $D$-term contribution that is comparable to the $F$-term contribution. 
The total masses depend in this case not only on the Goldstino direction of supersymmetry 
breaking, but also on the scalar VEV of the supersymmetry breaking fields. As for heterotic 
models, they can be flavor universal without any tuning only if both the modular weights 
$n_\alpha$ and the $U(1)$ charges $q_{\alpha a}$ are universal. On the other hand, in 
the geometric moduli domination limit, it is enough that certain linear combinations of the 
latter are universal. A second possibility is that only the $S$ and $T_r$ supersymmetry 
breaking moduli are stabilized by $F$-term effects, whereas the $U_s$ moduli 
behave as a minimal set of Higgs fields with a logarithmic K\"ahler potential and are stabilized 
by the $D$-terms in such a way to approximately compensate them. In this radically different 
situation, the resulting soft scalar masses have an accidentally vanishing $D$-term contribution 
and are thus entirely controlled by the $F$-term contribution. The total masses depend also in 
this case on both the Goldstino direction of supersymmetry breaking and the scalar VEV of the 
supersymmetry breaking fields. The interesting novelty arising in this situation is that these 
masses can be flavor universal without any tuning under the mild constraint that these effective 
modular weights $n_\alpha$ are universal, independently of the $U(1)$ charges $q_{\alpha a}$. 
On the other hand, the Goldstino direction turns out to be constrained by gauge invariance. 
In the geometric moduli domination limit, no further simplification occurs in the soft masses, but
the $T_r$ moduli dominate over the $U_s$ moduli.

We should finally emphasized that we have restricted our attention to situations involving a minimal 
set of Higgs fields, which do at the same time the jobs of compensating the $D$-term potential and 
making the superpotential of the moduli gauge invariant. In this case, the effects of the Higgs fields 
can be entirely related to those of the moduli fields; as a consequence, the soft masses can then be 
parametrized in terms of the arbitrary $F$ terms of the sole moduli multiplets. One may however consider 
also more general non-minimal situations, where two different sets of Higgs fields are used respectively 
to compensate the $D$-term potential and to make the superpotential of the moduli gauge invariant. 
In that case, the effects of the Higgs fields will not be entirely determined by those of the moduli fields, 
and the extra Higgs multiplets will behave as additional fields of the hidden sector, together with the 
moduli fields; the soft masses will then depend on the arbitrary $F$ terms of not only the moduli but also 
the extra Higgs multiplets. In other words, whenever more charged fields than gauge symmetries
are involved in the hidden sector dynamics, the number of light gauge-invariant chiral multiplets
participating to supersymmetry breaking in the low-energy theory increases. One may then consider 
the particular limit in which these extra degrees of freedom decouple at the scale of supersymmetry 
breaking, for example because of a large mass preserving supersymmetry. In that case, one should 
then recover the same situation as for a minimal set of Higgs fields. This means that the minimal 
situation considered in this paper can actually be viewed as the most general situation compatible 
with the assumption that the hidden sector involves only the moduli.

\section*{Acknowledgments}

We thank R.~Rattazzi and A.~Uranga for many enlightening discussions. We are also grateful 
to B.~de~Carlos, E. Dudas, M.~Gomez--Reino and F.~Zwirner for useful comments. This work was 
supported by the Swiss National Science Foundation.


\small

\begin{thebibliography}{99}

\bibitem{gravmed1}
  R.~Barbieri, S.~Ferrara and C.~A.~Savoy,
  {\it Gauge models with spontaneously broken local supersymmetry},
  Phys.\ Lett.\  B {\bf 119} (1982) 343.

\bibitem{gravmed2}
  H.~P.~Nilles, M.~Srednicki and D.~Wyler,
  {\it Weak interaction breakdown induced by supergravity},
  Phys.\ Lett.\  B {\bf 120} (1983) 346.

\bibitem{gravmed3}
  L.~J.~Hall, J.~D.~Lykken and S.~Weinberg,
  {\it Supergravity as the messenger of supersymmetry breaking},
  Phys.\ Rev.\  D {\bf 27} (1983) 2359.
  
\bibitem{gravun1}
  A.~H.~Chamseddine, R.~Arnowitt and P.~Nath,
  {\it Locally supersymmetric grand unification},
  Phys.\ Rev.\ Lett.\  {\bf 49} (1982) 970.

\bibitem{gravun2}
  L.~E.~Ibanez,
  {\it Locally supersymmetric SU(5) grand unification},
  Phys.\ Lett.\  B {\bf 118} (1982) 73.
  
\bibitem{gravun3}
  N.~Ohta,
  {\it Grand unified theories based on local supersymmetry},
  Prog.\ Th.\ Phys.\  {\bf 70} (1983) 542.

\bibitem{KL}
   V.~S.~Kaplunovsky and J.~Louis,
   {\it Model independent analysis of soft terms in effective supergravity and in
   string theory},
   Phys.\ Lett.\ B {\bf 306} (1993) 269
   [hep-th/9303040].

\bibitem{BIM}
   A.~Brignole, L.~E.~Ibanez and C.~Munoz,
   {\it Towards a theory of soft terms for the supersymmetric Standard Model},
   Nucl.\ Phys.\ B {\bf 422} (1994) 125
   [Errat.\ B {\bf 436} (1995) 747]
   [hep-ph/9308271].
      
\bibitem{BIMS}
   A.~Brignole, L.~E.~Ibanez, C.~Munoz and C.~Scheich,
   {\it Some issues in soft SUSY breaking terms from dilaton / moduli sectors},
   Z.\ Phys.\ C {\bf 74} (1997) 157
   [hep-ph/9508258].

\bibitem{grs1}
  M.~Gomez-Reino and C.~A.~Scrucca,
  {\it Locally stable non-supersymmetric Minkowski vacua in supergravity},
  JHEP {\bf 0605} (2006) 015
  [hep-th/0602246].
 
\bibitem{grs2}
  M.~Gomez-Reino and C.~A.~Scrucca,
  {\it Constraints for the existence of flat and stable non-supersymmetric vacua
  in supergravity},
  JHEP {\bf 0609} (2006) 008
  [hep-th/0606273].
 
\bibitem{grs3}
  M.~Gomez-Reino and C.~A.~Scrucca,
  {\it Metastable supergravity vacua with F and D supersymmetry breaking}
  JHEP {\bf 0708} (2007) 091 
  [arXiv:0706.2785].

\bibitem{GS}
  M.~B.~Green and J.~H.~Schwarz,
  {\it Anomaly cancellation in supersymmetric D=10 gauge theory and superstring
  theory},
  Phys.\ Lett.\  B {\bf 149} (1984) 117.
 
\bibitem{4DGS}
  M.~Dine, N.~Seiberg and E.~Witten,
  {\it Fayet-Iliopoulos terms in string theory},
  Nucl.\ Phys.\  B {\bf 289} (1987) 589.
  
\bibitem{Drees}
  M.~Drees,
  {\it Intermediate Scale Symmetry Breaking and the Spectrum of Super Partners in
  Superstring Inspired Supergravity Models},
  Phys.\ Lett.\  B {\bf 181} (1986) 279.
  
\bibitem{HagelinKelley}
  J.~S.~Hagelin and S.~Kelley,
  {\it Sparticle masses as a probe of GUT physics},
  Nucl.\ Phys.\  B {\bf 342} (1990) 95.

\bibitem{BinetruyDudas}
  P.~Binetruy and E.~Dudas,
  {\it Gaugino condensation and the anomalous U(1)},
  Phys.\ Lett.\  B {\bf 389} (1996) 503
  [hep-th/9607172].
  
\bibitem{DvaliPomarol}
  G.~R.~Dvali and A.~Pomarol,
  {\it Anomalous U(1) as a mediator of supersymmetry breaking},
  Phys.\ Rev.\ Lett.\  {\bf 77} (1996) 3728
  [hep-ph/9607383].
   
\bibitem{Barreiroetall}
  T.~Barreiro, B.~de Carlos, J.~A.~Casas and J.~M.~Moreno,
  {\it Anomalous U(1), gaugino condensation and supergravity},
  Phys.\ Lett.\  B {\bf 445} (1998) 82
  [hep-ph/9808244].
  
\bibitem{Dudas1}
  E.~Dudas, S.~Pokorski and C.~A.~Savoy,
  {\it Soft scalar masses in supergravity with horizontal $U(1)_X$ gauge symmetry},
  Phys.\ Lett.\ B {\bf 369} (1996) 255
  [hep-ph/9509410].

\bibitem{Dudas2}
  E.~Dudas, C.~Grojean, S.~Pokorski and C.~A.~Savoy,
  {\it Abelian flavour symmetries in supersymmetric models},
  Nucl.\ Phys.\ B {\bf 481} (1996) 85
  [hep-ph/9606383].
  
\bibitem{KawamuraD1}
  Y.~Kawamura and T.~Kobayashi,
  {\it Soft scalar masses in string models with anomalous $U(1)$ symmetry},
  Phys.\ Lett.\ B {\bf 375} (1996) 141,
  [Errat.\ B {\bf 388} (1996) 867]
  [hep-ph/9601365].

\bibitem{KawamuraD2}
  Y.~Kawamura and T.~Kobayashi,
  {\it Generic formula of soft scalar masses in string models},
  Phys.\ Rev.\ D {\bf 56} (1997) 3844
  [hep-ph/9608233].
 
\bibitem{KawamuraD3}
  T.~Higaki, Y.~Kawamura, T.~Kobayashi and H.~Nakano,
  {\it Anomalous U(1) D-term contribution in type I string models},
  Phys.\ Rev.\ D {\bf 69} (2004) 086004
  [hep-ph/0308110].

\bibitem{ADM}
  N.~Arkani-Hamed, M.~Dine and S.~P.~Martin,
  {\it Dynamical supersymmetry breaking in models with a Green--Schwarz mechanism},
  Phys.\ Lett.\ B {\bf 431} (1998) 329
  [hep-ph/9803432].
  
\bibitem{ChoiDFF}
  K.~Choi and K.~S.~Jeong,
  {\it Supersymmetry breaking and moduli stabilization with anomalous U(1) gauge symmetry},
  JHEP {\bf 0608} (2006) 007
  [hep-th/0605108].
  
\bibitem{sugra1}
  E.~Cremmer, B.~Julia, J.~Scherk, S.~Ferrara, L.~Girardello and P.~van Nieuwenhuizen,
  {\it Spontaneous symmetry breaking and higgs effect in supergravity without
  cosmological constant},
  Nucl.\ Phys.\  B {\bf 147} (1979) 105.

\bibitem{sugra2}
  E.~Witten and J.~Bagger,
  {\it Quantization of Newton's constant in certain supergravity theories},
  Phys.\ Lett.\  B {\bf 115} (1982) 202.

\bibitem{sugra3}
  E.~Cremmer, S.~Ferrara, L.~Girardello and A.~Van Proeyen,
  {\it Coupling supersymmetric Yang-Mills theories to supergravity},
  Phys.\ Lett.\  B {\bf 116} (1982) 231.

\bibitem{sugra4}
  E.~Cremmer, S.~Ferrara, L.~Girardello and A.~Van Proeyen,
  {\it Yang-Mills theories with local supersymmetry: Lagrangian, transformation
  laws and superhiggs effect},
  Nucl.\ Phys.\  B {\bf 212} (1983) 413.

\bibitem{sugra5}
  J.~A.~Bagger,
  {\it Coupling the gauge invariant supersymmetric nonlinear sigma model to
  supergravity},
  Nucl.\ Phys.\  B {\bf 211} (1983) 302.

\bibitem{FKZ}
   S.~Ferrara, C.~Kounnas and F.~Zwirner,
   {\it Mass formulae and natural hierarchy in string effective supergravities},
   Nucl.\ Phys.\ B {\bf 429} (1994) 589
   [Errat.\ B {\bf 433} (1995) 255]
   [hep-th/9405188].
  
\bibitem{dudasvempati}
  E.~Dudas and S.~K.~Vempati,
  {\it Large D-terms, hierarchical soft spectra and moduli stabilisation},
  Nucl.\ Phys.\  B {\bf 727} (2005) 139
  [hep-th/0506172].
  
\bibitem{KawamuraDFF}
  Y.~Kawamura,
  {\it Model independent analysis of soft masses in heterotic string models  with
  anomalous U(1) symmetry},
  Phys.\ Lett.\ B {\bf 446} (1999) 228
  [hep-ph/9811312].

\bibitem{noscale}
  J.~R.~Ellis, C.~Kounnas and D.~V.~Nanopoulos,
  {\it No Scale Supersymmetric Guts},
  Nucl.\ Phys.\  B {\bf 247} (1984) 373.

\bibitem{sequestered}
  L.~Randall and R.~Sundrum,
  {\it Out of this world supersymmetry breaking}
  Nucl.\ Phys.\  B {\bf 557} (1999) 79
  [hep-th/9810155].
  
\bibitem{PomarolDimopoulos}
  A.~Pomarol and S.~Dimopoulos,
  {\it Superfield derivation of the low-energy effective theory of softly broken
  supersymmetry},
  Nucl.\ Phys.\  B {\bf 453} (1995) 83
  [hep-ph/9505302].
  
\bibitem{Rattazzi}
  R.~Rattazzi,
  {\it A Note on the effective soft SUSY breaking Lagrangian below the GUT
  scale},
  Phys.\ Lett.\  B {\bf 375} (1996) 181
  [hep-ph/9507315].

\bibitem{STU1}
  E.~Witten,
  {\it Dimensional Reduction Of Superstring Models},
  Phys.\ Lett.\  B {\bf 155} (1985) 151.

\bibitem{STU2}
  S.~Ferrara, C.~Kounnas and M.~Porrati,
  {\it General dimensional reduction of ten-dimensional supergravity and
  superstring},
  Phys.\ Lett.\  B {\bf 181} (1986) 263.

\bibitem{STU3}
  M.~Cvetic, J.~Louis and B.~A.~Ovrut,
  {\it A string calculation of the Kahler potentials for moduli of Z(N)
  orbifolds},
  Phys.\ Lett.\  B {\bf 206} (1988) 227.

\bibitem{BdA}
  R.~Brustein and S.~P.~de Alwis,
  {\it Moduli potentials in string compactifications with fluxes: mapping the
  discretuum},
  Phys.\ Rev.\  D {\bf 69} (2004) 126006
  [hep-th/0402088].

\bibitem{HetOrb1}
  L.~J.~Dixon, J.~A.~Harvey, C.~Vafa and E.~Witten,
  {\it Strings on orbifolds},
  Nucl.\ Phys.\  B {\bf 261} (1985) 678.

\bibitem{HetOrb2}
  L.~J.~Dixon, J.~A.~Harvey, C.~Vafa and E.~Witten,
  {\it Strings on orbifolds. 2},
  Nucl.\ Phys.\  B {\bf 274} (1986) 285.
  
\bibitem{MetricHet}
  L.~J.~Dixon, V.~Kaplunovsky and J.~Louis,
  {\it On effective field theories describing (2,2) vacua of the heterotic
  string},
  Nucl.\ Phys.\  B {\bf 329} (1990) 27.
  
\bibitem{IbanezLust}
  L.~E.~Ibanez and D.~Lust,
  {\it Duality anomaly cancellation, minimal string unification and the effective
  low-energy Lagrangian of 4-D strings},
  Nucl.\ Phys.\  B {\bf 382} (1992) 305
  [hep-th/9202046].
       
\bibitem{InterBrane}
  G.~Aldazabal, S.~Franco, L.~E.~Ibanez, R.~Rabadan and A.~M.~Uranga,
  {\it Intersecting brane worlds},
  JHEP {\bf 0102} (2001) 047
  [hep-ph/0011132].
       
\bibitem{SusyInt1}
  M.~Cvetic, G.~Shiu and A.~M.~Uranga,
  {\it Three-family supersymmetric standard like models from intersecting  brane
  worlds},
  Phys.\ Rev.\ Lett.\  {\bf 87} (2001) 201801
  [hep-th/0107143].
  
\bibitem{SusyInt2}
  M.~Cvetic, G.~Shiu and A.~M.~Uranga,
  {\it Chiral four-dimensional N = 1 supersymmetric type IIA orientifolds from
  intersecting D6-branes},
  Nucl.\ Phys.\  B {\bf 615} (2001) 3
  [hep-th/0107166].
         
\bibitem{Cremades}
  D.~Cremades, L.~E.~Ibanez and F.~Marchesano,
  {\it SUSY quivers, intersecting branes and the modest hierarchy problem},
  JHEP {\bf 0207} (2002) 009
  [hep-th/0201205].

\bibitem{VZ}
  G.~Villadoro and F.~Zwirner,
  {\it D terms from D-branes, gauge invariance and moduli stabilization in  flux
  compactifications},
  JHEP {\bf 0603} (2006) 087
  [hep-th/0602120].
  
\bibitem{MetricInt1}
  D.~Lust, P.~Mayr, R.~Richter and S.~Stieberger,
  {\it Scattering of gauge, matter, and moduli fields from intersecting branes},
  Nucl.\ Phys.\  B {\bf 696} (2004) 205
  hep-th/0404134].
  
\bibitem{MetricInt2}
  M.~Bertolini, M.~Billo, A.~Lerda, J.~F.~Morales and R.~Russo,
  {\it Brane world effective actions for D-branes with fluxes},
  Nucl.\ Phys.\  B {\bf 743} (2006) 1
  [hep-th/0512067].
  
\bibitem{ReviewInt}
  R.~Blumenhagen, B.~Kors, D.~Lust and S.~Stieberger,
  {\it Four-dimensional string compactifications with D-Branes, orientifolds  and
  fluxes},
  Phys.\ Rept.\  {\bf 445} (2007) 1
  [hep-th/0610327].

\bibitem{SoftInt1}
  D.~Lust, S.~Reffert and S.~Stieberger,
  {\it MSSM with soft SUSY breaking terms from D7-branes with fluxes},
  Nucl.\ Phys.\  B {\bf 727} (2005) 264
  [hep-th/0410074].
  
\bibitem{SoftInt2}
  A.~Font and L.~E.~Ibanez,
  {\it SUSY-breaking soft terms in a MSSM magnetized D7-brane model},
  JHEP {\bf 0503} (2005) 040
  [hep-th/0412150].
           
\end{thebibliography}
\end{document}